\documentclass[fleqn,usenatbib]{mnras}

\usepackage[T1]{fontenc}
\DeclareRobustCommand{\VAN}[3]{#2}
\let\VANthebibliography\thebibliography
\def\thebibliography{\DeclareRobustCommand{\VAN}[3]{##3}\VANthebibliography}

\usepackage{graphicx}	
\usepackage{amsmath}	
\usepackage{amssymb}	
\usepackage[flushleft]{threeparttable}
\usepackage{hyperref}

\newcommand{\mbh}{\ensuremath{M_{\rm BH}}}
\newcommand{\mdot}{\ensuremath{\dot{m}/\dot{m}_{\rm Edd}}}
\newcommand{\rg}{\ensuremath{r_{\rm g}}}
\newcommand{\ltransf}{\ensuremath{L_{\rm transf}/L_{\rm disc}}}
\newcommand{\fcol}{\ensuremath{f_{\rm col}}}

\newcommand{\xmm}{\textit{XMM-Newton}}
\newcommand{\nustar}{\textit{NuSTAR}}
\newcommand{\swift}{\textit{Swift}}

\usepackage{amsmath} 
\let\oldAA\AA
\renewcommand{\AA}{\text{\normalfont\oldAA}}

\title[The first X-ray look at the most luminous quasar in the last 9\,Gyr]{The first X-ray look at SMSS\,J114447.77-430859.3: the most luminous quasar in the last 9\,Gyr}

\author[E. S. Kammoun et al.]{E. S. Kammoun,$^{1,2}$\thanks{E-mail: \url{ekammoun@irap.omp.eu}}
Z. Igo,$^{3}$
J. M. Miller,$^{4}$
A. C. Fabian,$^{5}$
M. T. Reynolds,$^{4,6}$
A. Merloni,$^{3}$
D. Barret,$^{1}$
\newauthor
E. Nardini,$^{2}$
P. O. Petrucci,$^{7}$
E. Piconcelli,$^{8}$
S. Barnier,$^{7}$
J. Buchner,$^{3}$
T. Dwelly,$^{3}$
I. Grotova,$^{3}$
\newauthor
M. Krumpe,$^{9}$
T. Liu,$^{3}$
K. Nandra,$^{3}$
A. Rau,$^{3}$
M. Salvato,$^{3}$
T. Urrutia,$^{9}$
J. Wolf $^{3}$
\\
$^{1}$ IRAP, Universit\'{e} de Toulouse, CNRS, UPS, CNES, 9, Avenue du Colonel Roche, BP 44346, F-31028, Toulouse Cedex 4, France \\
$^{2}$ INAF -- Osservatorio Astrofisico di Arcetri, Largo Enrico Fermi 5, I-50125 Firenze, Italy\\
$^{3}$ Max-Planck-Institut für Extraterrestrische Physik (MPE), Giessenbachstrasse 1, 85748 Garching bei München, Germany\\
$^{4}$ Department of Astronomy, The University of Michigan, 1085 S. University Ave., Ann Arbor, MI 48109, USA;\\
$^{5}$ Institute of Astronomy, University of Cambridge, Madingley Road, Cambridge CB3 OHA, UK\\
$^{6}$ Department of Astronomy, The Ohio State University, 140 West 18th Avenue, Columbus, OH, 43210, USA\\
$^{7}$ Univ. Grenoble Alpes, CNRS, IPAG, 38000 Grenoble, France\\
$^{8}$ INAF, Osservatorio Astronomico di Roma, Via Frascati 33, I– 00078 Monte Porzio Catone, Italy\\
$^{9}$ Leibniz-Institut für Astrophysik, Potsdam (AIP), An der Sternwarte 16, 14482 Potsdam, Germany
}
\date{Accepted XXX. Received YYY; in original form ZZZ}

\pubyear{2022}

\begin{document}
\label{firstpage}
\pagerange{\pageref{firstpage}--\pageref{lastpage}}
\maketitle

\begin{abstract}
SMSS\,J114447.77-430859.3 ($z=0.83$) has been identified in the SkyMapper Southern Survey as the most luminous quasar in the last $\sim 9\,\rm Gyr$. In this paper, we report on the eROSITA/\textit{Spectrum-Roentgen-Gamma} (SRG) observations of the source from the eROSITA All Sky Survey, along with presenting results from recent monitoring performed using \swift, \xmm, and \nustar. The source shows a clear variability by factors of $\sim 10$ and $\sim 2.7$ over timescales of a year and of a few days, respectively. When fit with an absorbed power law plus high-energy cutoff, the X-ray spectra reveal a $\Gamma=2.2 \pm 0.2$ and $E_{\rm cut}=23^{+26}_{-5}\,\rm keV$. Assuming Comptonisation, we estimate a coronal optical depth and electron temperature of $\tau=2.5-5.3\, (5.2-8)$ and $kT=8-18\, (7.5-14)\,\rm keV$, respectively, for a slab (spherical) geometry. The broadband SED is successfully modelled by assuming either a standard accretion disc illuminated by a central X-ray source, or a thin disc with a slim disc emissivity profile. The former model results in a black hole mass estimate of the order of $10^{10}\,M_\odot$, slightly higher than prior optical estimates; meanwhile, the latter model suggests a lower mass. Both models suggest sub-Eddington accretion when assuming a spinning black hole, and a compact ($\sim 10\,r_{\rm g}$) X-ray corona. The measured intrinsic column density and the Eddington ratio strongly suggest the presence of an outflow driven by radiation pressure. This is also supported by variation of absorption by an order of magnitude over the period of $\sim 900\,\rm days$.

\end{abstract}
\begin{keywords}
accretion, accretion discs -- galaxies: active -- galaxies: nuclei -- (galaxies:) quasars: general -- quasars: supermassive black holes -- quasars: individual: SMSS\,J114447.77-430859.3 -- X-rays: general
\end{keywords}



\section{Introduction}
\label{sec:intro}

Active galactic nuclei (AGN) are thought to be powered by the accretion of matter onto a supermassive black hole (SMBH) in the form of a disc. The primary hard X-ray continuum in AGN arises from repeated Compton up-scattering of UV/soft X-ray accretion disc photons in a hot, trans-relativistic plasma. This process typically results in a power-law spectrum extending to energies determined by the electron temperature in the hot corona \citep[e.g.,][]{Light88, Haardt93}. Broad-band UV/X-ray spectra require that the corona does not fully cover the disc \citep{Haa94}. X-ray microlensing experiments are suggestive of a compact corona in some bright quasars (QSOs) with a half-light radius smaller than $\sim 6\,r_{\rm g}$ \citep[e.g.,][]{Char09, Mosquera13}, where $r_{\rm g} =G\mbh/c^2$ is the gravitational radius. Eclipses of the X-ray source have also placed constraints on the size of the hard X-ray emitting region(s): $r \leq 10^{14}$\,cm \citep[e.g.,][]{Risaliti07}. In addition, spectral-timing studies of X-ray reflection in AGN are also suggestive of a centrally located, compact corona \citep[e.g.,][]{Marinucci16, Kara16}.

Most of our detailed knowledge of AGN, and in particular of their X-ray properties, is based on the study of nearby, low-mass, low-accretion rate sources. However, understanding the black hole growth, the energetics of AGN, and the disc-corona connection (among other properties) would require deep observations of high-mass and high-accretion rate sources. Accretion theory predicts that at high accretion rates, the standard radiatively efficient, optically thick and geometrically thin disc \citep{Shak73} breaks down. In this case, the radiation pressure becomes more important, and the disc becomes both optically and geometrically thick, a configuration known as a slim disc \citep[e.g.,][]{Abramowicz88}. However, recent studies showed that the broadband spectral energy distributions of super- and sub-Eddington AGN do not seem to exhibit any strong differences \citep[e.g.,][]{Castello-Mor17, Liu21}. Whilst high-mass sources are expected to show little or no variability, surprisingly, X-ray observations showed that $\sim 15-25\%$ of these AGN  exhibit large variability (exceeding a factor of 10) in the super-Eddington regime \citep[ e.g.,][]{Liu19,Liu21}. Moreover, a sizable fraction  ($\sim 30-40\%$) of high-Eddington quasars seems to be X-ray weak, both at intermediate ($z = 0.5-1$) and high ($z= 2-4$) redshift \citep[e.g.,][]{Nardini19,Zappacosta20,Laurenti22}. This fraction is much higher than the fraction of X-ray weak sources in non-jetted ``standard'' QSOs.

It is worth noting that the studies of such sources have a strong impact on our understanding black hole growth, as well as their impact on their close environment. In fact, the formation of SMBHs with masses of the order of $M_{\rm BH} > 10^9\,M_\odot$ at redshifts of $z \sim 6-7$ (i.e. when the Universe was less than 1\,Gyr old) is still debated. It has been proposed that one of the channels of forming such massive black holes would be via gas accretion at rates comparable/higher than the Eddington limit \citep{Johnson2016}. In addition, these sources are thought to launch powerful nuclear outflows \citep[][]{Nardini15, Matzeu17}, imprinting absorption lines in the X-ray spectra (especially in the $6-8$\,keV range), that are capable of regulating the growth and the evolution of their host galaxies \citep[e.g.,][]{King15,Giustini19}. Thus, highly accreting sources are unique laboratories to study AGN feedback, probing the real impact of nuclear activity on the evolution of massive galaxies and on the formation of structures in the Universe.

\subsection*{SMSS\,J114447.77-430859.3}

\begin{figure*}
\centering
\includegraphics[width=0.95\linewidth]{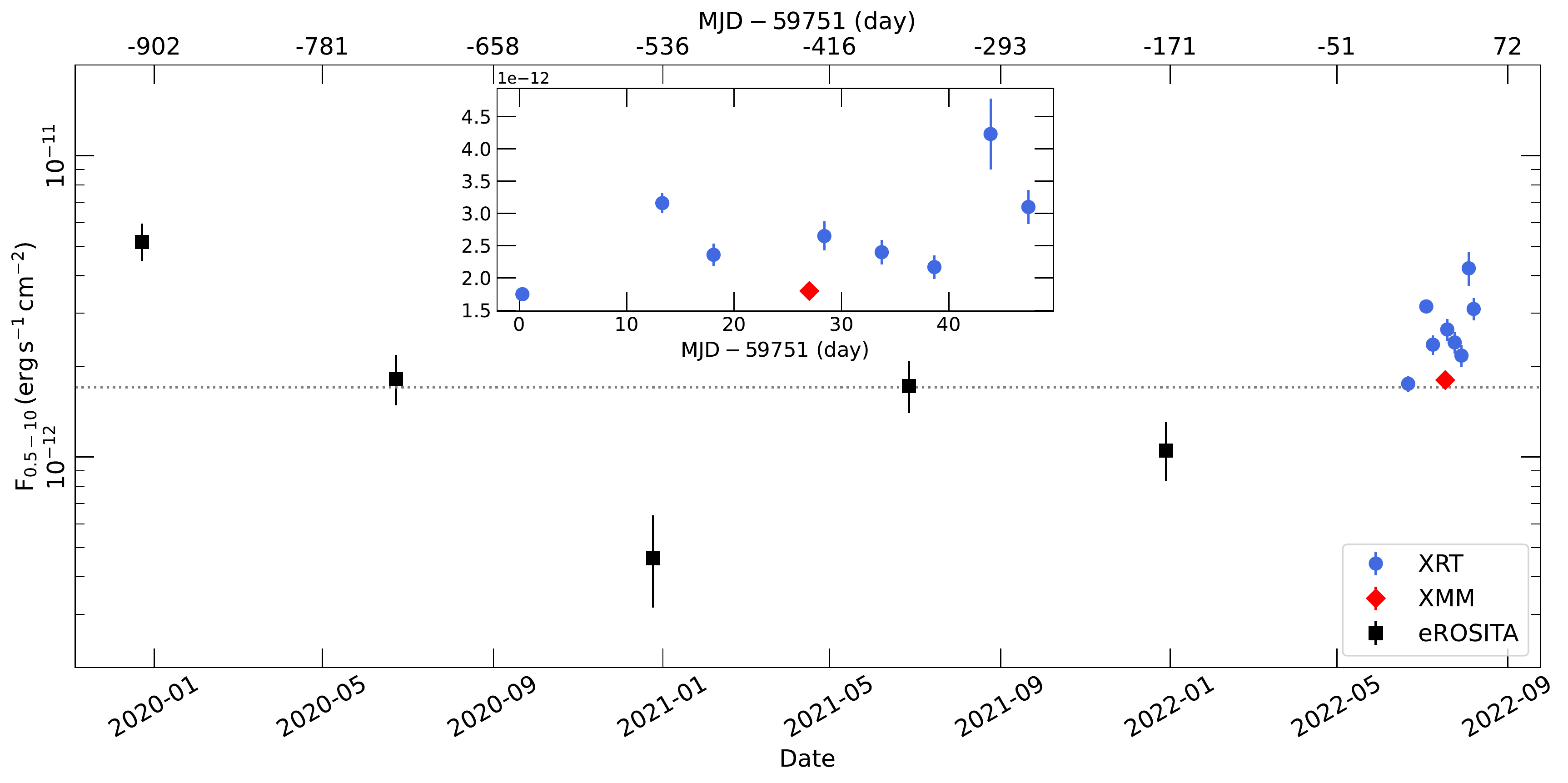}
\caption{The light curve of J1144 in the 0.5-10\,keV band. The black squares, blue circles, and red diamond correspond to eROSITA, \swift, and \xmm, respectively. The inset shows a zoom-in on the recent monitoring of the source in 2022, with the start being the first \swift\ observation ($\rm MJD = 59751$). The grey dotted line corresponds to the $3\sigma$ upper detection limit by ROSAT obtained in 1990 (see text for details).}
\label{fig:xraylc}
\end{figure*}

\cite{Onken22} reported on the discovery of SMSS\,J114447.77-430859.3 ($z=0.83$; hereafter, J1144) in the SkyMapper Southern Survey \citep[SMSS;][]{Wolf18,Onken19}, as the most luminous quasar of the last $\sim9$\,Gyr observed so far. Using broad hydrogen and magnesium emission lines, the authors estimated a black hole mass of $\log \left( M_{\rm BH}/M_\odot \right) = 9.4 \pm 0.5$. They also estimated the bolometric luminosity to be $L_{\rm bol} = (4.7 \pm 1.0) \times 10^{47}\, \rm erg\,s^{-1}$, suggesting an Eddington ratio of $\lambda_{\rm Edd} = L_{\rm bol}/L_{\rm Edd} \simeq 1.5_{-1.1}^{+3.3}$. This source was not detected in the ROSAT All Sky Survey (RASS). However, interestingly, the examination of five single passes of the eROSITA All-Sky Survey (eRASS1--5), which are each around 4 times deeper than RASS, confirmed not only the detection of this source, but also a large variability over the course of two years as seen in Fig.\,\ref{fig:xraylc}. In this paper, we present the eRASS1--5 observations of this source. We also present the results obtained from our recent monitoring of the source using the \textit{Neil Gehrels Swift Observatory (Swift)}, \textit{XMM-Newton}, and \textit{NuSTAR}, performed in 2022. J1144 is a unique, closer proxy of luminous QSOs, usually found at cosmic noon ($z\sim 2-3$), that require very deep observations in order to study their X-ray/UV/optical properties.

 The paper is structured as follows: the data reduction is presented in Section\,\ref{sec:data reduction}. Section\,\ref{sec:Xray} shows the results obtained by analysing the X-ray spectra. The broadband SED is presented in Section\,\ref{sec:SED}. Finaly, we discuss our results in Section\,\ref{sec:discussion}.
 
\section{Observations and data reduction}
\label{sec:data reduction}

\subsection{eROSITA observations}
\label{sec:erosita}
The extended ROentgen Survey with an Imaging Telescope Array \citep[eROSITA;][]{Predehl2021}, the soft X-ray instrument on board the \textit{Spektrum-Roentgen-Gamma (SRG)} orbital observatory, observed J1144 over a period of two years as it performed its first five eRASS passes (eRASS1--5). The eRASS Ecliptic scanning strategy means that an object at intermediate Ecliptic latitudes (like J1144) passes in the eROSITA field of view around $8-10$ times every 6 months, with those passes (each less than 40s) all occurring within 1--2 days. 

Data were reduced in the standard way using the eROSITA Science Analysis Software System (eSASS) \mbox{eSASSusers\_211214} pipeline version c020 and c947 for eRASS1--4 and 5, respectively \citep[][Merloni et al., in prep.]{Brunner2022}. A simple astrometric match was used to identify the eROSITA-detected source, as this is robust for high count rate sources with sub-arcsecond positional uncertainties. Exposure corrected spectral extraction of a circular source and annular background region was carried out using $\texttt{srctool}$ as described in Section 2.2 of \citet{TengLiu2022}. The source extraction radius was defined in a way to maximize the signal to noise and increases in size with the flux of the source, whilst the inner radius of the background annulus is chosen such that the surface brightness of the source's PSF wings is $<2\%$ of the local background surface brightness; both regions mask out other contaminating sources \citep{Brunner2022, TengLiu2022}. Subsequently, spectral fitting was done using the \texttt{pyXspec} X-ray analysis environment of {\tt XSPEC v12.12.0} \citep{Arnaud96, pyxspec} coupled with Bayesian X-ray Analysis \citep[\texttt{BXA v4.0.0};][]{Buchner14,Buchner21}, a Bayesian parameter estimation and model comparison software using the nested sampling algorithm UltraNest \citep{UltranestBuchner}. The spectra, grouped by each single-pass eRASS, from the source and background regions were jointly fitted using a source model, typically an absorbed power law, plus background model. The background model was calculated following \cite{Simmonds18}, by applying Principal Component Analysis (PCA) on an existing parametric model for eROSITA background spectra and adding Gaussian lines until it no longer improved the fit, as judged by the Akaike Information Criterion \citep[see also Section 3.1 of][for more details]{TengLiu2022}. The shape of this component was therefore already fixed in the joint modelling, however, the normalisation was left free to vary so it could adjust to the required background flux level. 

Light curves were also extracted using eSASS and analysed using $\texttt{bexvar}$ to search for variability \citep{Buchner2022}. Variability in $\texttt{bexvar}$ is quantified by the intrinsic scatter ($\sigma_{\mathrm{bexvar}}$) on the assumed log-normal distribution of count rates in any given time bin \citep[see Sec.\,3 of][]{Buchner2022}. This log-scatter on the log-count rate is similar in concept to the excess variance on the linear count rate, usually quoted as the normalized excess variance \citep[NEV; e.g.,][]{Vaughan2003}.

\begin{table}
\centering
\caption{Log of the X-ray observations. The net count rates are reported in the $0.2-5\rm\,keV$ range for eROSITA, the $0.4-5\rm\,keV$ for XRT, $0.4-10\,\rm keV$ for \xmm, and $3.5-30\,\rm keV$ for \nustar.}
\label{tab:Xrayobs}
\begin{tabular}{lllr} 

\hline \hline									
Date	&	Instrument	&	Net Count rate			&	Net exp.	\\
(MJD)	&		&	($\rm Count\,s^{-1}$)			&	(ks)	\\ \hline
58840	&	eROSITA/eRASS1	&$	3.7\pm0.3 $&	0.15 	\\
59023	&	eROSITA/eRASS2	&$	1.5\pm0.3	$&	0.10	\\
59208	&	eROSITA/eRASS3	&$	0.4\pm0.2	$&	0.10	\\
59392	&	eROSITA/eRASS4	&$	1.2\pm0.3	$&	0.11	\\
59577	&	eROSITA/eRASS5	&$	1.0\pm0.3	$&	0.12	\\ \hline
59751	&	XRT/O1	&$	0.039	\pm	0.003	$&	3.6	\\
59764	&	XRT/O3	&$	0.059	\pm	0.005	$&	2.2	\\
59769	&	XRT/O4	&$	0.041	\pm	0.005	$&	1.7	\\
59779	&	XRT/O6	&$	0.042	\pm	0.005	$&	1.4	\\
59785	&	XRT/O7	&$	0.036	\pm	0.005	$&	1.7	\\
59790	&	XRT/O8	&$	0.037	\pm	0.004	$&	1.6	\\
59795	&	XRT/O9	&$	0.056	\pm	0.008	$&	0.7	\\
59798	&	XRT/O10	&$	0.058	\pm	0.007	$&	1.2	\\ \hline
59778	&	\xmm/PN	&$	0.538	\pm	0.007	$&	9.7	\\
	&	\xmm/MOS2	&$	0.158	\pm	0.003	$&	13.7	\\ \hline
59759	&	\nustar/FPMA	&$	0.024	\pm	0.0006	$&	61.6	\\
	&	\nustar/FPMB	&$	0.022	\pm	0.0007	$&	60.9	\\ \hline

\end{tabular}

\end{table}

\subsection{\swift\ observations}
\label{sec:swift}

\swift\ observed J1144 in 10 occasions over a period of 47\,days, from 2022-06-21 to 2022-08-07 (Obsid 15227001-15227010, hereafter XRT/O1--O10, respectively). Two of these observations (O2 and O5) were too short, resulting in less than 20 counts in total, thus we omitted them from this analysis. The X-ray telescope \citep[XRT;][]{Burrows05} operated during these observations in the Photon Counting (PC) mode. We reduced the data following standard procedures using {\tt HEASOFT} \citep{heasoft}. We performed the initial reduction with {\tt xrtpipeline}. Source and background spectra were extracted using {\tt xselect} from circular regions of 50\arcsec\ in radius.  We used the default redistribution matrix file (RMF) and ancillary response file (ARF), available in the calibration database. The spectra were then binned, requiring a minimum signal-to-noise ratio (S/N) of 3 per energy bin. 

\begin{figure}
\centering
\includegraphics[width=0.95\linewidth]{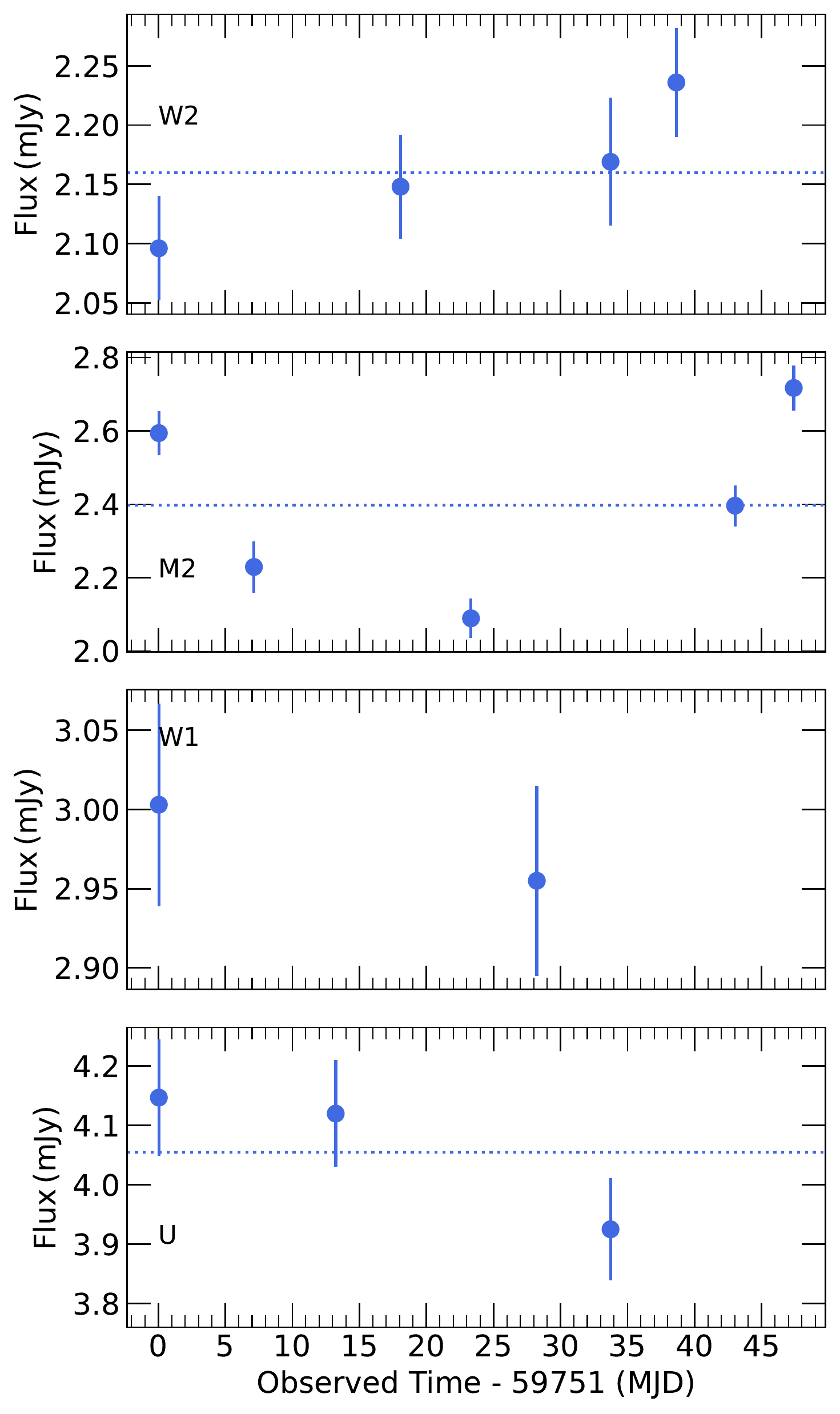}
\caption{UVOT light curves obtained during the monitoring of the source in 2022. The horizontal dotted lines represent the average flux in the bands with more than two data points.}
\label{fig:optlightcurve}
\end{figure}

J1144 was simultaneously observed in the optical/UV by the \swift/UVOT \citep{Roming05}. All observations were analyzed in the {\tt HEASOFT-6.30.1} environment utilizing the latest version of the UVOT \textsc{caldb}. All image segments contained in a given obsid are summed prior to source detection. The flux from J1144 was calculated using the \texttt{uvotmaghist} task with a 5\arcsec\ source region centred on the known coordinates of J1144. Background was estimated from a nearby source free region. The UVOT light curves are shown in Fig.\,\ref{fig:optlightcurve}.

\subsection{\xmm\ observation}
\label{sec:xmm}

\xmm\ observed J1144 on 2022-07-18 (ObsID 0911791701) for a total exposure of 15\,ks. The observation was operated in the Full Frame/Thin Filter mode for EPIC-pn \citep{Stru01} and the two EPIC-MOS \citep{Tur01} instruments. We reduced the data using SAS v.19.1.1 \citep{Gabriel04} and the latest calibration files. We processed the data using {\tt EPPROC} and {\tt EMPROC} for the EPIC-pn and EPIC-MOS, respectively. Source spectra were extracted from a circular region of radius $30\arcsec$ centered on the source. The corresponding background spectra were extracted from an off-source circular region located on the same CCD chip, with a radius twice that of the source. We filtered out periods with strong background flares estimated to be around 5.5\,ks for EPIC-pn and 1.4\,ks for EPIC-MOS.  Response matrices were produced using the {\tt FTOOLs} {\tt RMFGEN} and {\tt ARFGEN}. We rebinned the observed spectra using the {\tt SAS} task {\tt SPECGROUP} to have a minimum S/N of 4 in each energy bin. The EPIC-MOS1 observation suffered from a bad column coincident with the location of the source. Thus, we use in the following analysis the results from EPIC-pn and EPIC-MOS2 only.

\subsection{\nustar\ observation}
\label{sec:nustar}

J1144 was observed by \nustar\ on 2022-06-28 (ObsID 90801616002) for a total exposure of 130\,ks (net exposure of $\sim 61$\,ks).  The data were reduced using the standard pipeline in the \nustar\ Data Analysis Software (NUSTARDAS v2.0.0), and using the latest calibration files. We cleaned the unfiltered event files with the standard depth correction. We reprocessed the data using the ${\tt saamode = optimized}$ and ${\tt tentacle = yes}$ criteria for a more conservative treatment of the high background levels in the proximity of the South Atlantic Anomaly. We extracted the source and background light curves and spectra from circular regions of radii $50\arcsec$, for the two focal plane modules (FPMA and FPMB) using the {\tt HEASOFT} task {\tt nuproducts}. We binned the spectra to require a minimum S/N of 4 in each energy bin. The \nustar\ light curves do not show any signature of variability during the observation. In the following we analyse the spectra from FPMA and FPMB jointly, without combining them together.

The details of each of the observations are shown in Table\,\ref{tab:Xrayobs}. Figure\,\ref{fig:xraylc} shows the observed flux in the $0.5-10$\,keV light curves for eROSITA, \swift, and \xmm\ (black squares, blue circles, and red diamond, respectively). The fluxes are estimated based on the best-fit models (see next section for details). We used the HIgh-energy LIght curve GeneraTor\footnote{\url{http://xmmuls.esac.esa.int/hiligt/}} \citep[HILIGT;][]{Saxton22} to derive an upper limit on the ROSAT non-detection. This gives a $3\sigma$ upper limit on the $0.2-2\rm\,keV$ flux of $1.4\times 10^{-12}\,\rm erg\,s^{-1}\,cm^{-2}$. Assuming $\Gamma = 2$, this translates into a $0.5-10\,\rm keV$ flux upper limit of $1.7\times 10^{-12}\, \rm erg\,s^{-1}\,cm^{-2}$. This is consistent with the flux values during eRASS2--5 and the early observations of the 2022 monitoring.

\begin{table}
\centering
\caption{UVOT light curve of the source obtained in the differnt filters.}
\label{tab:UVobs}
\begin{tabular}{lll} 

\hline \hline									
Date	&	Filter	&	Flux 			\\
(MJD)	&		&			(mJy)			\\ \hline
59751	&	W2	&$	2.10	\pm	0.04	$\\
59751	&	M2	&$	2.59	\pm	0.06	$\\
59751	&	W1	&$	3.00	\pm	0.06	$\\
59751	&	U	&$	4.15	\pm	0.10	$\\
59751	&	B	&$	4.90	\pm	0.10	$\\
59751	&	V	&$	6.24	\pm	0.16	$\\
59758	&	M2	&$	2.23	\pm	0.07	$\\
59764	&	U	&$	4.12	\pm	0.09	$\\
59769	&	W2	&$	2.15	\pm	0.04	$\\
59774	&	M2	&$	2.09	\pm	0.05	$\\
59779	&	W1	&$	2.96	\pm	0.06	$\\
59785	&	U	&$	3.93	\pm	0.09	$\\
59785	&	W2	&$	2.17	\pm	0.05	$\\
59790	&	W2	&$	2.24	\pm	0.05	$\\
59794	&	M2	&$	2.40	\pm	0.06	$\\
59798	&	M2	&$	2.72	\pm	0.06	$\\ \hline

\end{tabular}

\end{table}

\begin{table*}
\centering

\caption{Best-fit parameters obtained by modelling the eROSITA spectra. Note that the instrinsic column density ($N_{\rm H}$) is a 3$\sigma$ upper limit. Normalisations (norm1--5) are in units of $\rm Photon\,keV^{-1}\,cm^{-2}\,s^{-1}$.}
\label{tab:erositaspec}
\begin{tabular}{ccccccc} 

\hline \hline																
$N_{\rm H} {[\rm cm^{-2}]}$	&	$\Gamma$							&	$\rm \log (norm1)$ &	$\rm \log (norm2)$ &	$\rm \log (norm3)$ &	$\rm \log (norm4)$ &	$\rm \log (norm5)$					\\ \hline
$< 2.7 \times 10^{21}$  &  $2.21_{-0.12}^{+0.14}$  & $-2.27_{-0.05}^{+0.06}$  & $-2.72_{-0.06}^{+0.07}$  & $-3.31_{-0.10}^{+0.09}$  & $-2.75_{-0.06}^{+0.06}$  & $-2.97_{-0.06}^{+0.08}$  \\ \hline
\end{tabular}
\end{table*}

\section{X-ray spectral analysis}
\label{sec:Xray}

\subsection{eRASS results}
\label{sec:eRASSspectra}

As mentioned before, we use $\texttt{bexvar}$ to assess whether any significant temporal variability exists within any given eRASS, by looking at the log-scatter on the log-count rate. For J1144, this typical `intra-eRASS' variability timescale is around 4 hours. However, low values of $\sigma_{\mathrm{bexvar}} \ll 0.1$ are consistently found for all eRASS1--5 over several energy bands. Overall, the long term (inter-eRASS) X-ray variability, clearly seen in Fig.\,\ref{fig:xraylc}, is much more significant than the short term (intra-eRASS) variability. For that reason, in the following, we model the time-averaged spectrum from each eRASS.

\begin{figure}
\centering
\includegraphics[width=0.99\linewidth]{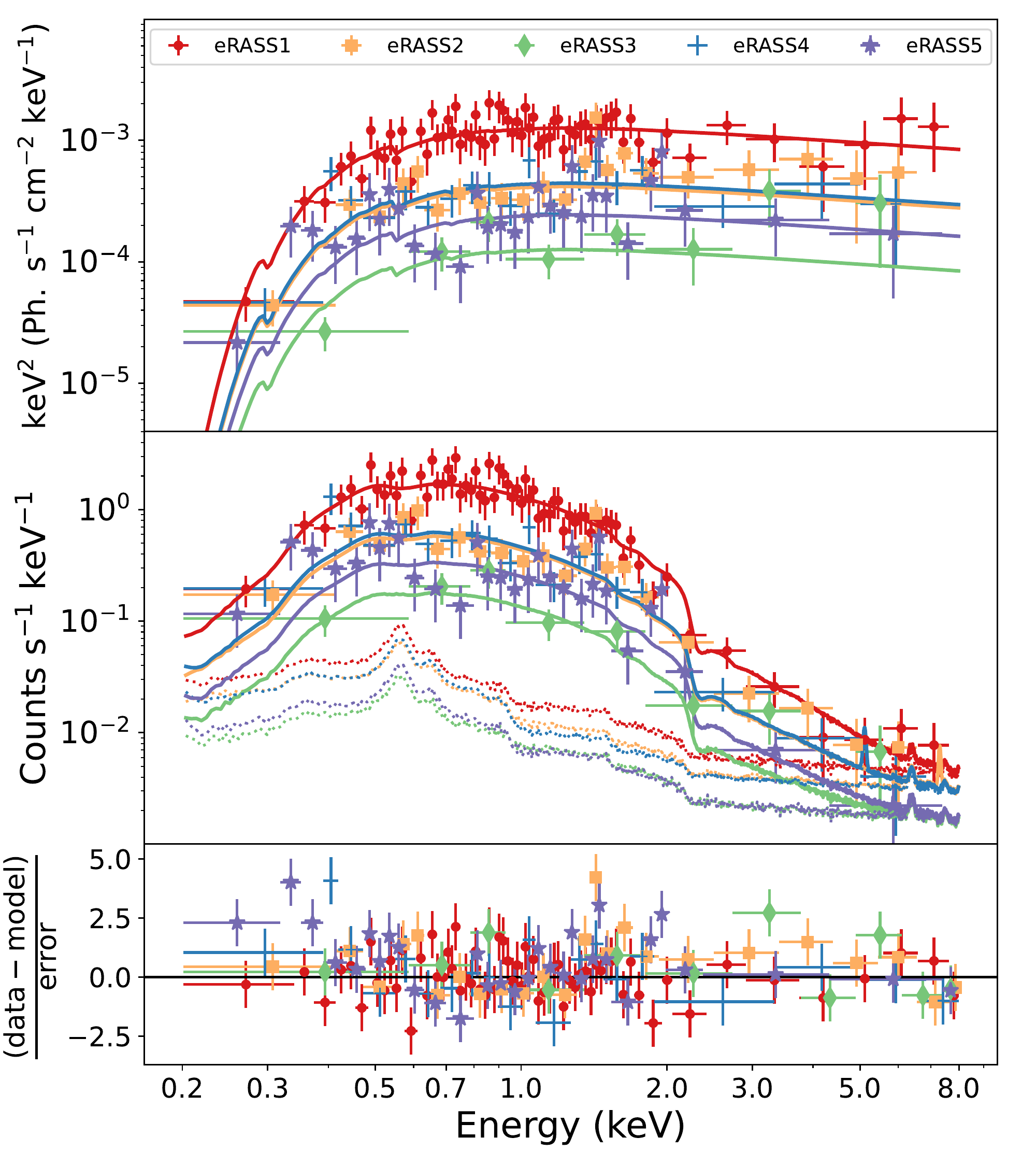}
\caption{Observed spectra from the different eROSITA observations (eRASS1$-$5: red circle, yellow square, green diamond, blue cross, purple star). Top: Deconvolved spectra. Middle: Convolved spectra, folded through detector response but not effective area, along with background spectra for each eRASS (dotted, same colour scheme). Bottom: normalized residuals obtained by fitting the spectra.}
\label{fig:erositaspec}
\end{figure}

We model the spectra using a simple absorbed power law ({${\tt TBabs \times zTBabs \times zpow}$ in XSPEC formalism), where {\tt TBabs} and {\tt zTBabs} \citep{Wilms00} correspond to the Galactic absorption in the line of sight and the intrinsic neutral absorption at the redshift of the source, respectively. We fixed the Galactic column density at $N_{\rm H}= 7.9\times 10^{20}\,\rm cm^{-2}$ \citep{nhPaper} but the intrinsic column density (“znH1--5”), the power law photon index (“PhoIndex1-5”), and power-law normalisations (“norm1-5”) were left free for each eRASS spectrum. Upon modelling each spectrum individually, it is found that the intrinsic column density and photon index are consistent with each other across eRASS1--5, albeit with large uncertainties, due to the quality of the data. Yet, the values of the normalisations were showing a clear variability across the five spectra. For this reason, we re-fit the spectra simultaneously by tying the column density and the photon index for all observations, and keeping the normalisation free to vary. The prior set on the column density, the photon index, and the normalisations was a log-uniform prior ranging from $(10^{-3}-10^{3})\times10^{22}\,\rm cm^{-2}$, a uniform prior from $1-4$, and a log-uniform prior from 10$^{-6}-10\, \rm Photon\, keV^{-1}\, cm^{-2} \,s^{-1}$, respectively. 

Figure \ref{fig:erositaspec} showcases the results of this simultaneous fitting of eRASS1--5 spectra. The top panel shows the deconvolved spectra, to better compare with observations from different instruments. The middle panel shows the convolved spectra, folded through detector response but not effective area, along with background spectra (dotted lines). It is clear that the background becomes dominant past $2-3$\,keV. The normalized residuals are also displayed in the bottom panel of the same figure. The apparent difference in the levels of background are caused by the varying extraction region sizes, which scale proportional to the source flux. Quantitatively, the eRASS1 (highest flux state) source extraction region area is a factor of $\sim$3 larger than that of eRASS3 (lowest flux state), and this is reflected in the difference in normalization between the red and green dotted lines. The best-fit (posterior median) photon index and 3$\sigma$ upper limit on the column density for eRASS1--5 are $\Gamma = 2.21^{+0.14}_{-0.12}$ and $N_{\rm H} < 2.7 \times 10^{21}~\rm{cm}^{-2}$, respectively. Moreover, we find a factor $>10$ decrease in the normalisation of the power law between eRASS1 and eRASS3, after which it steadily increases again up till eRASS5. Table\,\ref{tab:erositaspec} lists the best-fit parameters for this simultaneous fitting of eRASS1--5 spectra. The corner plot in Figure\,\ref{fig:cornererosita} shows the parameter space sampled during the fit procedure and the posterior distributions obtained for each parameter (znH, PhoIndex, norm1--5). The confidence contours are drawn at 68th and 95th percentiles. Note that BXA methodology is fully consistent with the Levenberg–Marquardt minimization algorithm used in {\tt XSPEC}, however, BXA is preferred for modelling eROSITA sources as the sampling remains unbiased even in the low count regimes.}

\begin{figure*}
\centering
\includegraphics[width=0.9\linewidth]{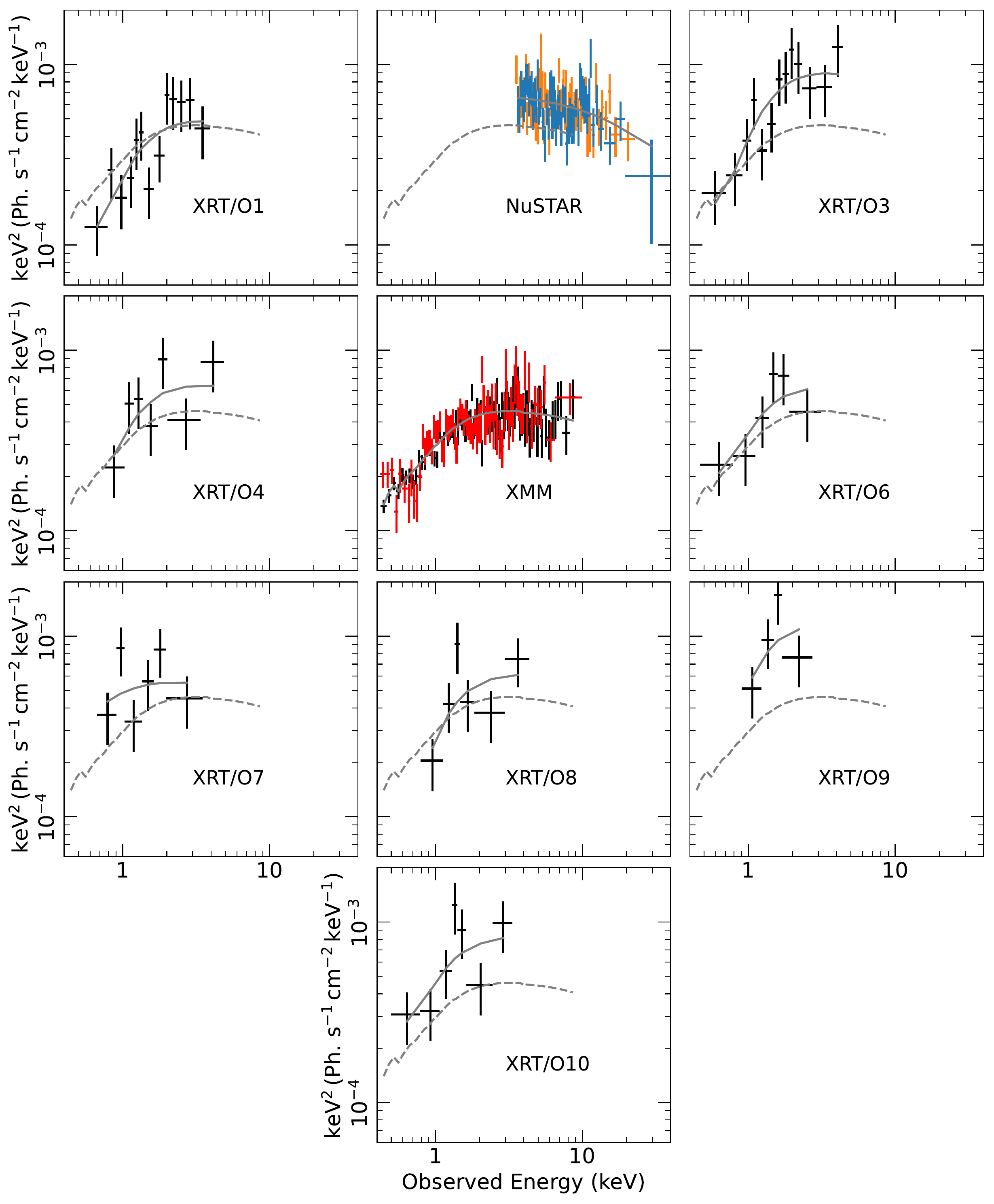}
\caption{\swift, \xmm, and \nustar\ spectra obtained during the monitoring of J1144 in 2022. The solid lines correspond to the best-fit model assuming an absorbed power law (see Sec.\,\ref{sec:monitoring}). The dashed lines represent the best-fit model to the \xmm\ data, for comparison.}
\label{fig:xrayspec}
\end{figure*}
\subsection{2022 monitoring}
\label{sec:monitoring}

\begin{figure}
\centering
\includegraphics[width=0.95\linewidth]{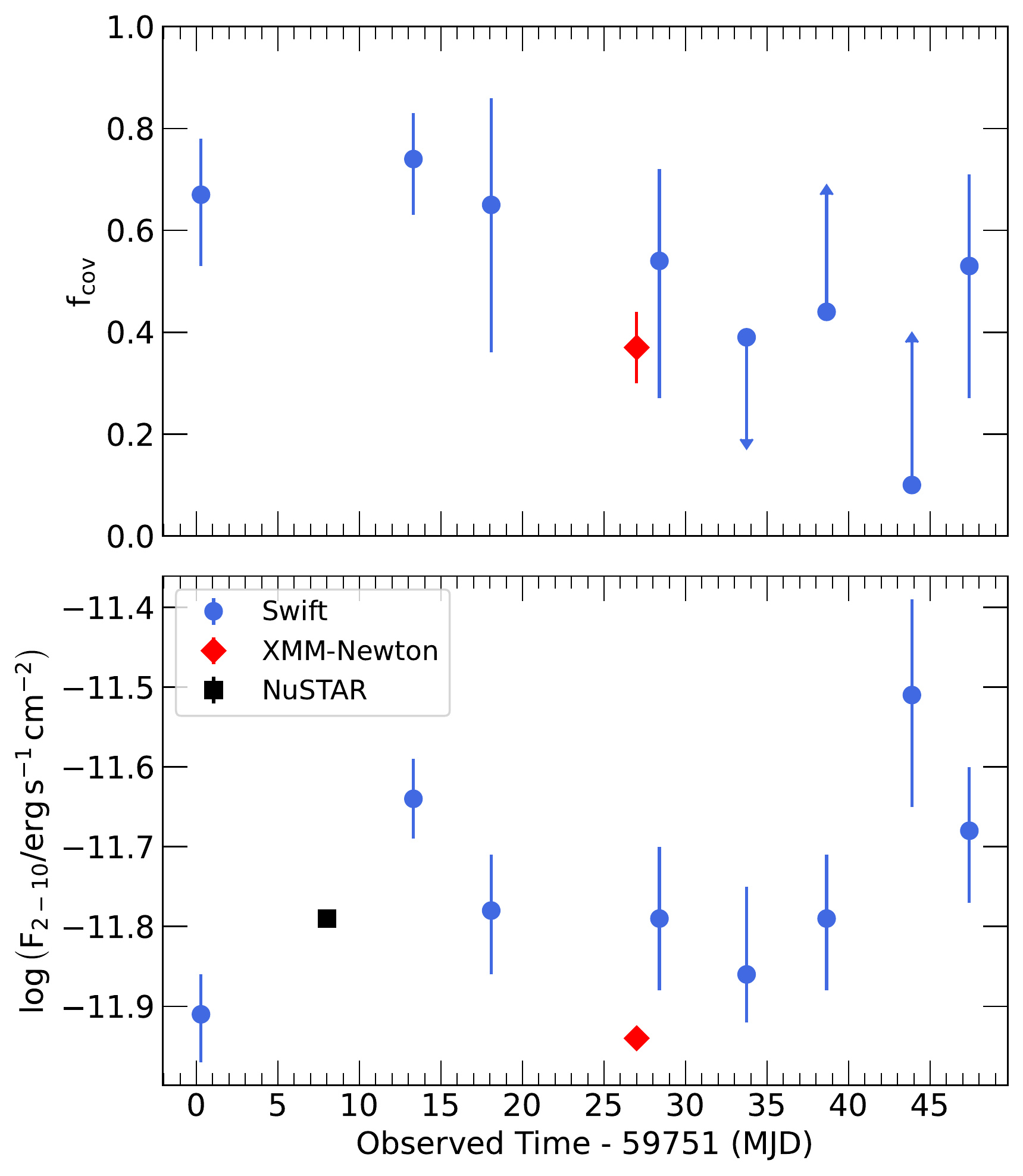}
\caption{Variability of $f_{\rm cov}$ and the $2-10$\,keV flux during the monitoring of J1144 in 2022.}
\label{fig:fcovlogF}
\end{figure}


In the rest of the spectral analysis we use \texttt{XSPEC v12.11.1} \citep{Arnaud96}. We fit simultaneously the data from all of the \swift/XRT, \xmm, and \nustar\ data assuming the following model:

\begin{equation}
    {\tt Model = TBabs \times TBpcf \times cflux \times zcutoffpl.}
    \label{eq:tbpcf}
\end{equation}

\noindent where {\tt TBabs} corresponds to the Galactic absorption in the line of sight of the source. {\tt TBpcf} corresponds to a partially covering neutral absorption at the rest frame of the source. The {\tt cflux} component measures the flux of the power-law component with a high-energy cutoff ({\tt zcutoffpl}) in the $2-10$\,keV range. We kept the column density of the {\tt TBpcf} component constant for all observations, but we let the covering fraction ($f_{\rm cov}$) free to vary. We also kept the photon index ($\Gamma$) and the high-energy cutoff ($E_{\rm cut}$) constant for all observations and we let the flux free. The spectra together with the best-fitted model are shown in Fig.\,\ref{fig:xrayspec}. The corresponding residuals are shown in Fig.\,\ref{fig:xrayresid}. All uncertainties are listed with $1\sigma$ confidence level. We present in Appendix\,\ref{sec:linescan} the results obtained by performing a line search on the \xmm\ spectra. This results in hints of an absorption line at $\sim 1.3\,\rm keV$, and an emission line at $\sim 6.5\,\rm keV$, with a significance of $\sim 3\sigma$.

The model is statistically accepted with $\rm \chi^2/dof = 346.7/320$ ($p_{\rm null} = 0.15$). We obtained a best-fit $\Gamma = 2.0 \pm 0.1$, $E_{\rm cut} = 68_{-23}^{+65}\, \rm keV$, and $N_{\rm H} = (3.7 \pm 0.6) \times 10^{22}\, \rm cm^{-2}$. The best-fit results show hints of variability in $f_{\rm cov}$ that cannot be strongly constrained due to the quality of the data. However, the changes in the $2-10$\,keV flux can be clearly confirmed with a max-to-min ratio of $\sim 2.7$ over the period of the monitoring (see Fig.\,\ref{fig:fcovlogF}). The best-fit photon index obtained from this monitoring of the source is consistent within uncertainties with the one obtained by modeling the eROSITA spectra.


For completeness, we tested a model by replacing {\tt TBpcf} in Eq.\,(\ref{eq:tbpcf}) by a warm absorption model, {\tt zxipcf}. We kept the column density and the covering fraction constant for all observations and let the ionisation parameter vary. This resulted in a worse fit than using the neutral absorption ($\rm \chi^2/dof = 357/320$). Keeping the ionisation parameter constant and letting the covering fraction free to vary results in a comparable fit to the one with {\tt TBpcf}, with an ionisation parameter consistent with a neutral absorption, deriving an upper limit of $\log \xi < 1.5$.

\begin{table}
\centering

\caption{Best-fit $f_{\rm cov}$ and $2-10$\,keV flux obtained by fitting the X-ray spectra of the source from the monitoring in 2022.}
\label{tab:Spec}
\begin{tabular}{lll} 

\hline \hline																
Observation		&	$f_{\rm cov}$							&	$\log (F_{2-10}/\rm cgs)$					\\ \hline
XRT/O1		&$	0.68	_{	-0.13	}^{+	0.05			}$ & $	-11.91	_{	-0.06	}^{+	0.05	}$ \\ [0.2cm]
XRT/O3		&$	0.75	_{	-0.11	}^{+	0.09			}$ & $	-11.64	_{	-0.05	}^{+	0.05	}$ \\ [0.2cm]
XRT/O4		&$	0.65	_{	-0.29	}^{+	0.21			}$ & $	-11.78	_{	-0.08	}^{+	0.07	}$ \\ [0.2cm]
XRT/O6		&$	0.56	_{	-0.27	}^{+	0.18			}$ & $	-11.79	_{	-0.09	}^{+	0.09	}$ \\ [0.2cm]
XRT/O7		&$	< 0.39							$ & $	-11.86	_{	-0.06	}^{+	0.11	}$ \\ [0.2cm]
XRT/O8		&$	>0.43$ & $	-11.81	_{	-0.09	}^{+	0.08	}$ \\ [0.2cm]
XRT/O9		&$	> 0.10							$ & $	-11.52	_{	-0.14	}^{+	0.12	}$ \\ [0.2cm]
XRT/O10		&$	0.55	_{	-0.26	}^{+	0.18			}$ & $	-11.68	_{	-0.09	}^{+	0.08	}$ \\ [0.2cm]
\xmm		&$	0.44	_{	-0.07	}^{+	0.07			}$ & $	-11.93	_{	-0.01	}^{+	0.01	}$ \\ [0.2cm]
\nustar		&$	-							$ & $	-11.79	_{	-0.01	}^{+	0.01	}$ \\ \hline
\end{tabular}
\end{table}

\section{Broadband spectral energy distribution}
\label{sec:SED}

\subsection{KYNSED}
\label{sec:kynsed}

\begin{figure*}
\centering
\includegraphics[width=0.49\linewidth]{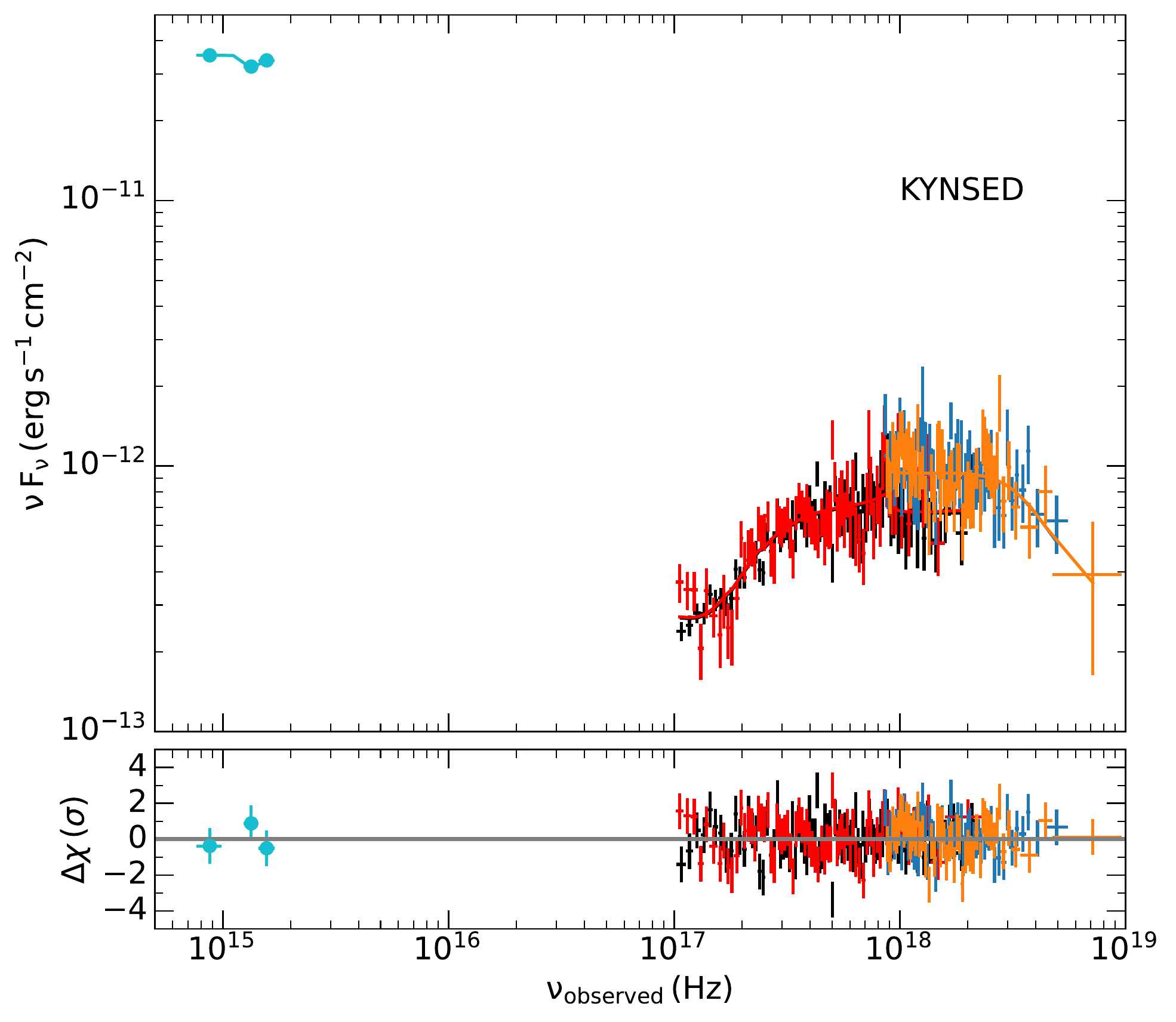}
\includegraphics[width=0.49\linewidth]{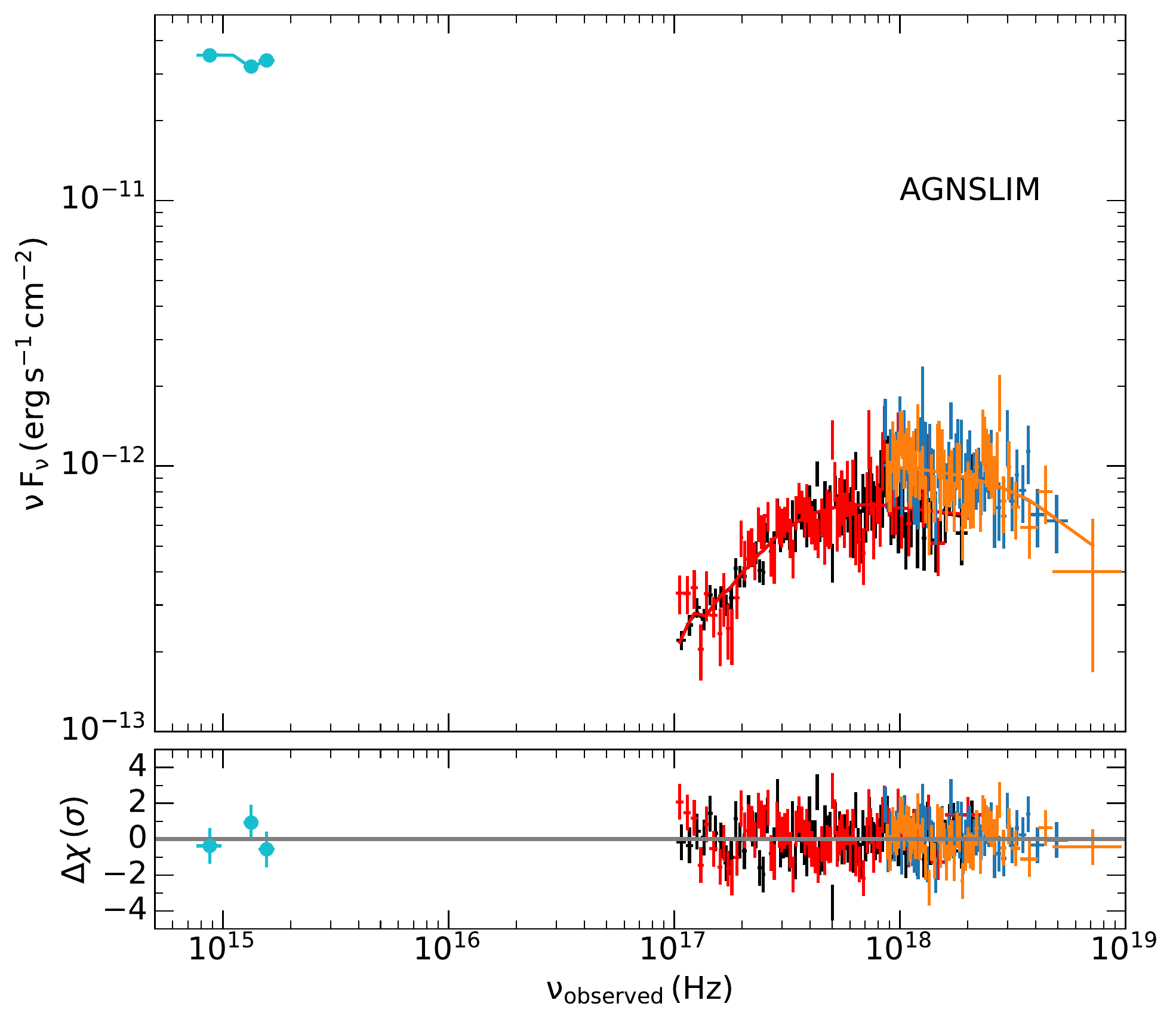}
\caption{Observed SED of J1144 using \swift/UVOT (cyan circles), \xmm/pn,MOS (black and red, respectively), and \nustar/FPMA,FPMB (orange and blue, respectively). The solid line corresponds to one of the best-fit models obtained for $a^\ast =0$ using {\tt KYNSED} (left) and {\tt AGNSLIM} (right). The other realisations resulted in a similar quality of the fit (see  text for details). The bottom panels show the residuals corresponding to these fits.}
\label{fig:SED}
\end{figure*}

First, we use {\tt KYNSED}\footnote{\url{https://projects.asu.cas.cz/dovciak/kynsed}} \citep{Dovciak22} to model the broadband SED of the source. This model considers a Novikov-Thorne \citep{Novikov73} accretion disc that powers the X-ray corona, assumed to be a point source located on the rotational axis of the SMBH. The fraction of the power transferred from the accretion disc to the corona is denoted by $\ltransf$. The X-ray corona emits then a primary power-law continuum which partially arrives to the observer and partially irradiates the disc. The disc will reprocess the incident emission. A part of this emission is re-emitted in the form of an X-ray reflection spectrum, and the other part is absorbed by the disc. This absorbed radiation will heat the disc and will be re-emitted in the UV/optical as an additional thermal emission. This process explained successfully the observed UV/optical continuum time-lags \citep[see e.g.,][]{Kammoun19,Kammoun21} as well as the power spectral density \citep{Panagiotou20,Panagiotou22}, obtained from intense monitoring of local AGN. \cite{Dovciak22} used {\tt KYNSED} to model the broadband SED obtained from the long monitoring of NGC\,5548. The authors discuss in their Section 5.5 the fact that the code does not take variability into account in its computations. Thus, the code is not ideal for modelling simultaneous X-ray/UV/optical observations. Instead, it is more suited for modeling time-averaged SEDs. For that reason, we constructed the time-averaged spectrum from the UVOT monitoring. To do so, we estimated the average flux of the source in bands with three or more observations. These are the UVW2, UVM2, and U bands, only. As for the X-ray spectra, we considered the \xmm\ and \nustar\ ones, which, together, should be representative of the average state of J1144.

In {\tt XSPEC} parlance, the model can be written as follows:
\begin{multline}
   {\tt Model = redden_{UV/opt} \times TBabs_{X} \times TBpcf_{X} \times KYNSED} 
    \label{eq:kynsed}
\end{multline}
\noindent All the relevant {\tt KYNSED} parameters are tied in the UV and optical range with those in the X-ray range. The ${\tt TBabs_X}$ component accounts for the effects of the Galactic absorption in the X-ray band. We fix its column density to $7.91\times 10^{20}\,\rm cm^{-2}$. The ${\tt redden_{UV/opt}}$ component accounts for the reddening due to dust extinction in our Galaxy. We fix the extinction $E(B-V)$ to 0.111 \citep{Schlafly11}. The assumed values of Galactic $N_{\rm H}$ and $E(B-V)$ are consistent with the observed relation between these two values \citep[e.g.,][]{bohlin78}. ${\tt TBpcf_X}$ accounts for partially covering absorption in the rest frame of the source. We applied ${\tt TBabs_X}$ and  ${\tt TBpcf_X}$ to the X-ray spectra only, while ${\tt redden_{UV/opt}}$ is only applied to the UV/optical data.

\begin{table}
\centering

\caption{Best-fit parameters obtained by modelling the SED of J1144 using {\tt KYNSED}, assuming different values of the spin. The uncertainties represent the lowest/highest values obtained from the SED modelling (see text for details).}
\label{tab:SED}
\begin{tabular}{lccc} 

\hline \hline																						
	&		$a^\ast = 0$					&			$a^\ast = 0.7$				&			$a^\ast = 0.998$				\\ \hline
$h\,(r_{\rm g})$	&	\multicolumn{3}{c}{	$	3-30	$	}	 \\[0.1cm]
$\theta (^\circ)$	&	\multicolumn{3}{c}{	$	0-60	$	}	\\[0.1cm]
$f_{\rm col}$	&	\multicolumn{3}{c}{	$	1.0-2.4	$	}	\\[0.1cm]
$\log \mbh/M_\odot$	&	$10.10_{-0.47}^{+0.38}$	&	$10.33_{-0.56}^{+0.43}$	&	$10.55_{-0.51}^{+0.62}$	\\[0.1cm]
$\log \dot{m}/\dot{m}_{\rm Edd}$	&	$0.19_{-0.28}^{+0.41}$	&	$0.00_{-0.34}^{+0.47}$	&	$-0.03_{-0.66}^{+0.54}$	\\[0.1cm]
$\Gamma$	&	$2.16_{-0.16}^{+0.25}$	&	$2.17_{-0.15}^{+0.27}$	&	$2.19_{-0.19}^{+0.28}$	\\[0.1cm]
$E_{\rm cut}\,\rm (keV)$	&	$23_{-5}^{+26}$	&	$22_{-6}^{+23}$	&	$22_{-7}^{+15}$	\\[0.1cm]
$L_{\rm transf, X}/L_{\rm disc}$	&	$0.08_{-0.04}^{+0.05}$	&	$0.07_{-0.03}^{+0.09}$	&	$0.05_{-0.02}^{+0.07}$	\\[0.1cm]
$L_{\rm transf, N}/L_{\rm disc}$	&	$0.11_{-0.05}^{+0.08}$	&	$0.10_{-0.05}^{+0.15}$	&	$0.07_{-0.03}^{+0.11}$	\\[0.1cm]
$N_{\rm H}\,\rm(10^{22}\,cm^{-2})$	&	$3.39_{-0.40}^{+0.46}$	&	$3.35_{-0.32}^{+0.52}$	&	$3.32_{-0.46}^{+0.51}$	\\[0.1cm]
$f_{\rm cov}$	&	$0.54_{-0.11}^{+0.12}$	&	$0.55_{-0.10}^{+0.13}$	&	$0.55_{-0.12}^{+0.13}$	\\[0.1cm] \hline
$\xi_{\rm in} \,(\rm erg\,cm\,s^{-1})$	&	$0.2_{-0.2}^{+2.7}$	&	$0.2_{-0.1}^{+4.8}$	&	$0.5_{-0.5}^{+63.5}$	\\[0.1cm]
$R_{\rm c}\,(r_{\rm g})$	&	$10.1_{-4.7}^{+14.5}$	&	$6.9_{-3.8}^{+19.3}$	&	$5.6_{-3.9}^{+12.1}$	\\[0.1cm] 
$\mathcal{R}$ & $0.75 _{- 0.25} ^{+ 0.23}$ & $0.83 _{- 0.24} ^{+ 0.27}$ & $0.89 _{- 0.25} ^{+ 0.68}$ \\ \hline
$\chi^2$	&	$272-294$	&	$271-300$	&	$271-300$	\\[0.1cm]							
dof	&	\multicolumn{3}{c}{	$	276	$	}	\\ \hline
\end{tabular}
\end{table}

The model does not give a statistically acceptable fit when we fix $M_{\rm BH}$ to the value that is reported by \cite{Onken22}. Instead, the fit improves significantly by letting the mass to be larger. Given that many parameters could not be constrained and that some of them are degenerate we adopted the following fitting scheme. We fix the spin to 0, 0.7, and 0.998. For each of the spin values, we choose randomly, assuming a uniform distribution, a combination of coronal height ($h$), colour correction factor\footnote{This factor corrects for the spectral hardening due to photon interactions with matter in the upper layers of the accretion disc.} ($f_{\rm col}$), and inclination ($\theta$). The limits of each parameter are given in Table\,\ref{tab:SED}. We fixed the parameters to the chosen values and we fit the SED letting the mass and the accretion rate in units of Eddington\footnote{The Eddington accretion rate is defined as $\dot{m}_{\rm Edd} = L_{\rm Edd}/\eta c^2$, where $\eta$, the radiative efficiency is a function of the BH spin, and $L_{\rm Edd} \simeq 1.26 \times 10^{38}\mbh/M_\odot\,\rm erg\,s^{-1}$ is the Eddington luminosity.} (\mdot) to be free. In addition, $\Gamma$, $E_{\rm cut}$, $N_{\rm H}$, and $f_{\rm cov}$ are left free but tied between the different spectra. We left $L_{\rm transf}/L_{\rm disc}$ untied between \xmm\ and \nustar\ ($L_{\rm transf,X}/L_{\rm disc}$ ab=nd $L_{\rm transf,N}/L_{\rm disc}$, respectively). We repeated this for 500 times for each of the spin values. We selected only the fits that are statistically accepted with ($\rm \chi^2/dof < 300/276$, i.e., $p_{\rm null}> 0.15 $).

Assuming Comptonisation, and using the conservation of photons during this process, {\tt KYNSED} estimates the size of the corona ($R_{\rm c}$) a posteriori \citep[see][for more details]{Dovciak16, Ursini20, Dovciak22}. This assumes a spherical source on the rotation axis of the BH. Despite the fact that this computation is approximate, it proved to be aligned with more accurate 3D models \citep[see Sec. 5.3 in][]{Dovciak22}. We use this approximation to rule out configurations in which the best fit results in a large corona that is not consistent with the assumptions of the model. This can be used to make an additional selection on the accepted configurations. For that reason, we select from our fits only the results where $R_{\rm c}$ is smaller than the difference between the height of the source and the event horizon of the BH. This results in a selection of 80\,\% of the realisations that are statistically accepted and physically consistent with the assumptions of the model.

\begin{figure*}
\centering
\includegraphics[width=0.49\linewidth]{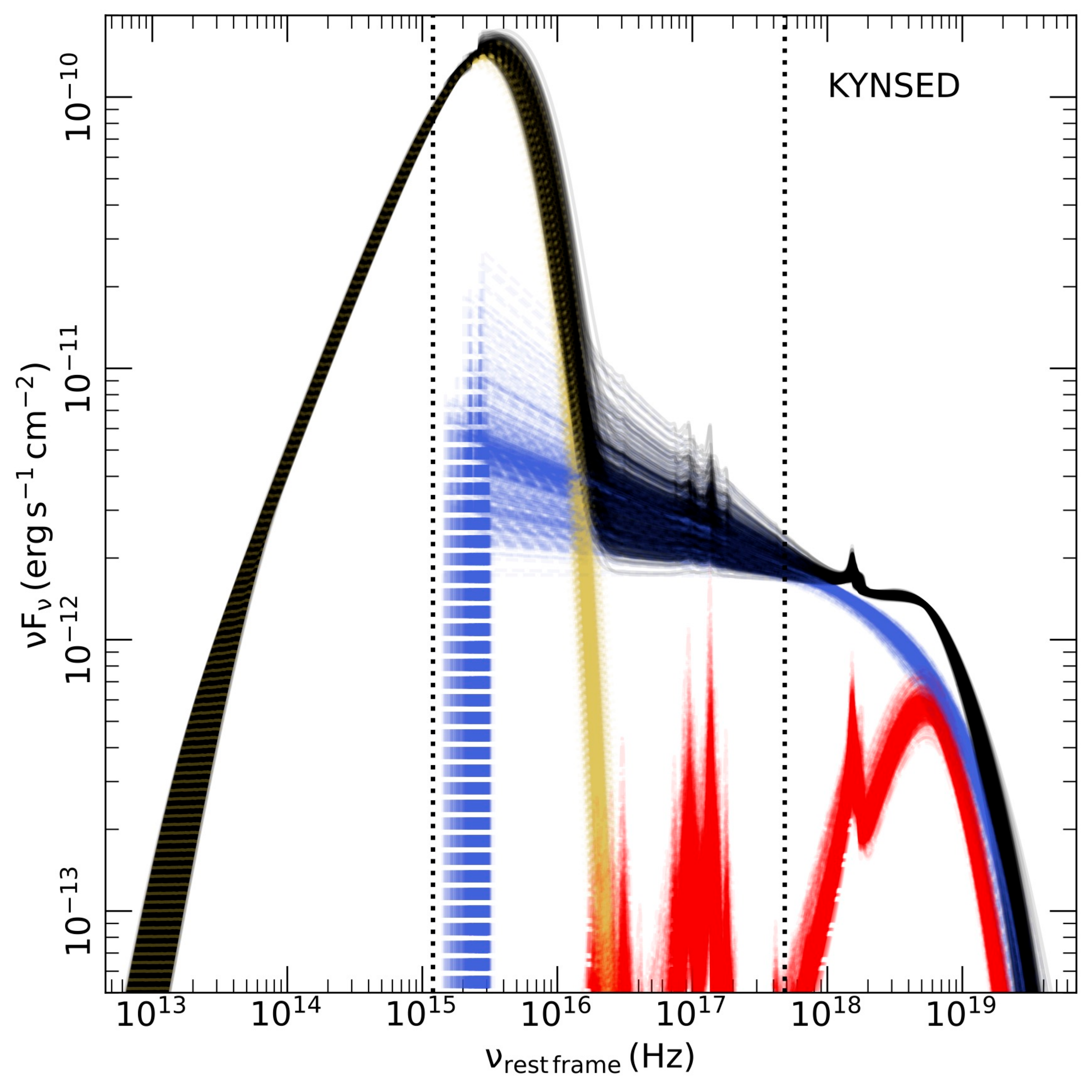}
\includegraphics[width=0.49\linewidth]{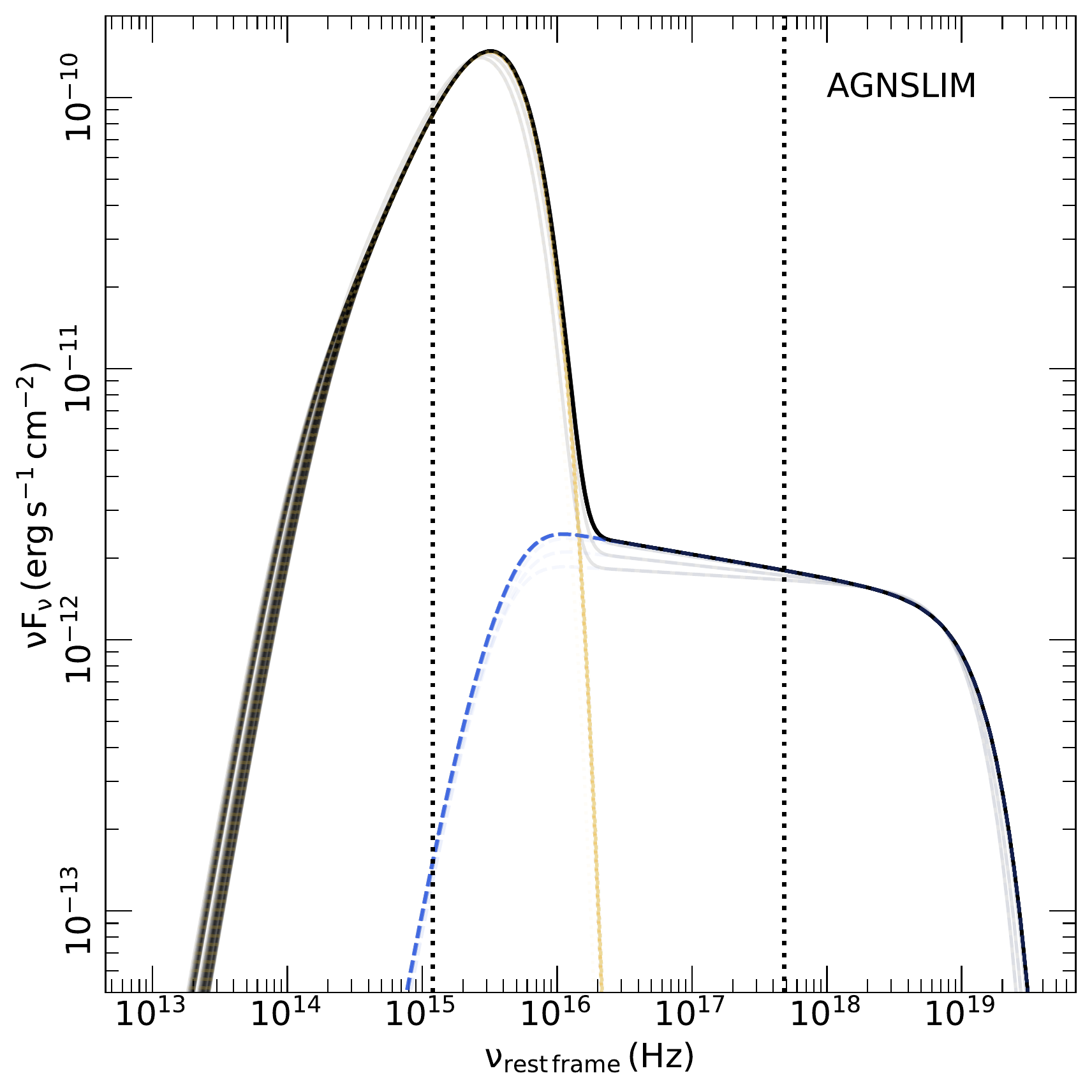}

\caption{Model SEDs (in the rest frame of the source) obtained from all of the realisations using {\tt KYNSED} (left) and {\tt AGNSLIM} (right). The total model is shown in black. The thermal emission of the disc is shown in yellow. The power law component is shown using the dashed blue lines. The dotted red lines correspond to the X-ray reflection from {\tt KYNSED} ({\tt AGNSLIM} does not take this component into account).  The dotted vertical lines correspond to the intrinsic $2500\,\AA$ and 2\,keV.}
\label{fig:SEDmodel}
\end{figure*}

Figure\,\ref{fig:SED} shows one realisation of the selected fits. The other realisations result in comparable fit quality. We show in Table\,\ref{tab:SED} the median value for each of the parameters. The uncertainties shown in this table correspond to the minimum/maximum value obtained for each of the parameters obtained from the 500 fits. We also show in this table the minimum-maximum values of $\chi^2$ for each spin. All values are statistically accepted. Figure\,\ref{fig:SEDmodel} represents the unabsorbed SED models from all the realisations. We show in this figure the total SED in black, the power law component in blue, the reflection component in red, and the disc component in yellow.

All of the best-fit parameters are consistent for all spins, except the mass and the accretion rate which show a different behaviour. This can be seen in Fig.\,\ref{fig:cornerSED}, which shows all the best-fit parameters for each of the spin  values. The figure shows the expected degeneracy between $\Gamma$ and $E_{\rm cut}$. In addition it shows a degeneracy between \mbh, \mdot, and $f_{\rm col}$. The \mbh-\mdot\ degeneracy is better highlighted in Fig.\,\ref{fig:massmdot}. In this figure, we show each of the best-fit \mbh\ and \mdot\ for $a^\ast = 0, 0.7,$ and 0.998 (left, middle, and right panels, respectively). The fact that $\dot{m}$ in physical units is linked to \mdot\ by the efficiency ($\eta$), that is spin dependent, results in obtaining a lower \mdot\ range as the spin increases, for a given \mbh. For a given spin, changing $f_{\rm col}$ results in opposite behavior of \mbh\ and \mdot\ trying to compensate for the shift in the position of the model peak frequency. This leads to the observed degeneracy between \mdot, \mbh, and $f_{\rm col}$. The accretion rate changes between sub- and super-Eddington with $f_{\rm col}$. For $a^\ast = 0, 0.7,$ and 0.998, \mdot\ gets below the Eddington limit for $f_{\rm col}$ above 2.2, 1.8, and 1.6, respectively. For most of the possible combinations of parameters the needed \mbh\ value is larger than the one inferred from the width of the optical emission lines reported by \cite{Onken22}. The two values agree within uncertainties for low spin values and low $f_{\rm col}$. The discrepancy in mass is further discussed in Section\,\ref{sec:bhmass}.
 
Our SED modelling results in a bolometric luminosity that is consistent between all the assumed spins, $L_{\rm bol} = 6.2_{-0.5}^{+2.5} \times 10^{47}\,\rm erg\,s^{-1}$. This is higher than the value of $(4.7 \pm 1.0)\times 10^{47}\,\rm erg\,s^{-1}$ obtained by \cite{Onken22}, however, consistent within uncertainties. The measured values of $L_{\rm bol}$ and \mbh\ imply an Eddington ratio of $\lambda_{\rm Edd} = 0.34_{-0.18}^{+0.41},0.21_{-0.12}^{+0.26}, 0.11_{-0.06}^{+0.15}$, for $a^\ast = 0, 0.7, 0.998$, respectively. This is lower than the value of $1.5_{-1.1}^{+3.3}$ implied by \cite{Onken22}, albeit consistent within uncertainties.

\subsection{AGNSLIM}
\label{sec:agnslim}

As mentioned earlier, accretion discs in super-Eddington AGN are thought to follow a slim disc configuration. For that reason, in this section we fit the SED of J1144 using the {\tt AGNSLIM} model from \cite{Kubota19}. This model divides the disc in three regions: (a) an inner region (between $R_{\rm in}$ and $R_{\rm hot}$), where the luminosity is dissipated in hot slab-like material, forming a Comptonized spectrum, (b) an intermediate region (from $R_{\rm hot}$ to $R_{\rm warm}$) where the luminosity is dissipated in an optically thick and warm Comptonizing medium, and (c) the outer region (from $R_{\rm warm}$ to $R_{\rm out}$) that is completely thermal emitting a standard blackbody spectrum. This model assumes that the disc extends down to $R_{\rm in}$ with an emissivity following the one of a slim disc. It is worth noting that the slim disc is expected to give a large scale height. However, given the complexity of implementing such geometry, {\tt AGNSLIM} adopts a geometrically thin disc approximation, and it is limited to modifying the disc emissivity. Furthermore, the model does not include either reprocessing of hard X-rays or general relativity effects. We define the model in {\tt XSPEC} parlance as follows:
\begin{equation}
    {\tt Model = redden_{\rm UV/opt} \times TBabs_{\rm X} \times TBpcf_{\rm X} \times AGNSLIM}.
\end{equation}

\noindent Similarly to Eq.\,(\ref{eq:kynsed}), the ${\tt redden_{\rm UV/opt}}$ component represents the reddening due to Galactic absorption and is applied only to the UV/optical data. The ${\tt TBabs_{\rm X-ray}}$ and ${\tt TBpcf_{\rm X-ray}}$ components represent the Galactic and intrinsic absorption and they are applied only to the X-ray spectra. We fix $R_{\rm in}$ at its default value of $-1$, as calculated in Eq.\,(1) of \cite{Kubota19}. We assume the outer radius of the disc to be equal to the self-gravity radius as calculated by \cite{Laor89}. We fixed the spin values at 0, 0.7, and 0.998. For each spin value, we fixed cosine of the inclination ($\cos \theta$) at 20 values between 0.1 and 1. The free parameters are the BH mass, accretion rate, the temperature ($kT_{\rm h}$), the photon index ($\Gamma_{\rm h}$), and the radius ($R_{\rm h}$) of the hot corona, and the temperature ($kT_{\rm w}$), the photon index ($\Gamma_{\rm w}$), and the radius ($R_{\rm w}$) of the warm corona. The model resulted in statistically good fits, however the warm corona is not needed by the model as its radius always tends to be equal to the one of the hot corona. In fact, this result is expected as the warm corona is usually invoked to explain the presence of a soft X-ray excess (below $\sim 1\,\rm keV$), which is not seen in J1144. We repeated the fit by setting $R_{\rm w} = R_{\rm h}$, thus neglecting the presence of the warm corona. We linked all the parameters for the \xmm\ and \nustar\ spectra, except $R_{\rm h}$ which was left free to account for the difference in flux between the two observations. The fits are all acceptable with $\chi^2/\rm dof \simeq 277/276$, except for the high inclinations ($\cos \theta \leq 0.2$) of the non-spinning BH case. All of the good fits converged to the same value of $\Gamma_{\rm h}$, $kT_{\rm h}$, $N_{\rm H}$, and $f_{\rm cov}$ for all values of spin and inclination. The BH mass and the accretion rate varied with spin and inclination. However, the coronal size varied only as a function of spin. The best fit values are shown in Table\,\ref{tab:SED_agnslim}. The uncertainties on mass and mass accretion rate correspond to the lower and upper limits obtained by fitting the SED for different values of $\cos \theta$, for each spin value. The uncertainties on other parameters correspond to the statistical uncertainty on each of the parameter for a $\Delta \chi^2 = 1$. The right panel of Fig.\,\ref{fig:SED} shows the observed SED fitted with {\tt AGNSLIM} assuming $a^\ast = 0$. The right panel of Fig.\,\ref{fig:SEDmodel} shows the unabsorbed SED models from all the realisations. 

\begin{table}
\centering

\caption{Best-fit parameters obtained by modelling the SED of J1144 using {\tt AGNSLIM}, assuming different values of the spin. The uncertainties on mass and mass accretion rate represent the lowest/highest values obtained from the SED modelling (see text for details). The uncertainties on other parameters correspond to the statistical uncertainty on each of the parameter for a $\Delta \chi^2 = 1$.}
\label{tab:SED_agnslim}
\begin{tabular}{lccc} 

\hline \hline																						
	&		$a^\ast = 0$					&			$a^\ast = 0.7$				&			$a^\ast = 0.998$				\\ \hline
$\cos \theta$	&	$0.2-1$	&	$0.1-1$	&	$0.1-1$	\\[0.1cm]
$\log \dot{m}/\dot{m}_{\rm Edd}$	&	$0.18_{-0.10}^{+0.21}$	&	$-0.08_{-0.13}^{+0.37}$	&	$-0.70_{-0.13}^{+0.37}$	\\[0.1cm]
$\log \mbh/M_\odot$	&	$9.36_{-0.10}^{+0.20}$	&	$9.66_{-0.13}^{+0.37}$	&	$10.25_{-0.13}^{+0.37}$	\\[0.1cm]
$kT_{\rm h}$	&	$16_{-5}^{+14}$	&	$16_{-5}^{+14}$	&	$16_{-5}^{+14}$	\\[0.1cm]
$\Gamma_{\rm h}$	&	$2.09_{-0.04}^{+0.04}$	&	$2.09_{-0.04}^{+0.04}$	&	$2.09_{-0.04}^{+0.04}$	\\[0.1cm]
$R_{\rm h, X}$	&	$8.96_{-0.16}^{+0.19}$	&	$4.90_{-0.16}^{+0.19}$	&	$1.52_{-0.03}^{+0.03}$	\\[0.1cm]
$R_{\rm h, N}$	&	$9.58_{-0.26}^{+0.31}$	&	$5.22_{-0.31}^{+0.35}$	&	$1.58_{-0.03}^{+0.03}$	\\[0.1cm]
$N_{\rm H}\,\rm(10^{22}\,cm^{-2})$	&	$3.39_{-0.49}^{+0.53}$	&	$3.39_{-0.49}^{+0.53}$	&	$3.39_{-0.49}^{+0.53}$	\\[0.1cm]
$f_{\rm cov}$	&	$0.49^{+0.04}_{-0.03}$	&	$0.49^{+0.04}_{-0.03}$	&	$0.49^{+0.04}_{-0.03}$	\\[0.1cm]\hline
$\chi^2$	&	$277.3$	&	$277.3$	&	$277.5$	\\[0.1cm]
dof	&	\multicolumn{3}{c}{	$	276	$	}	\\ \hline
\end{tabular}
\end{table}

\begin{figure*}
\centering
\includegraphics[width=0.998\linewidth]{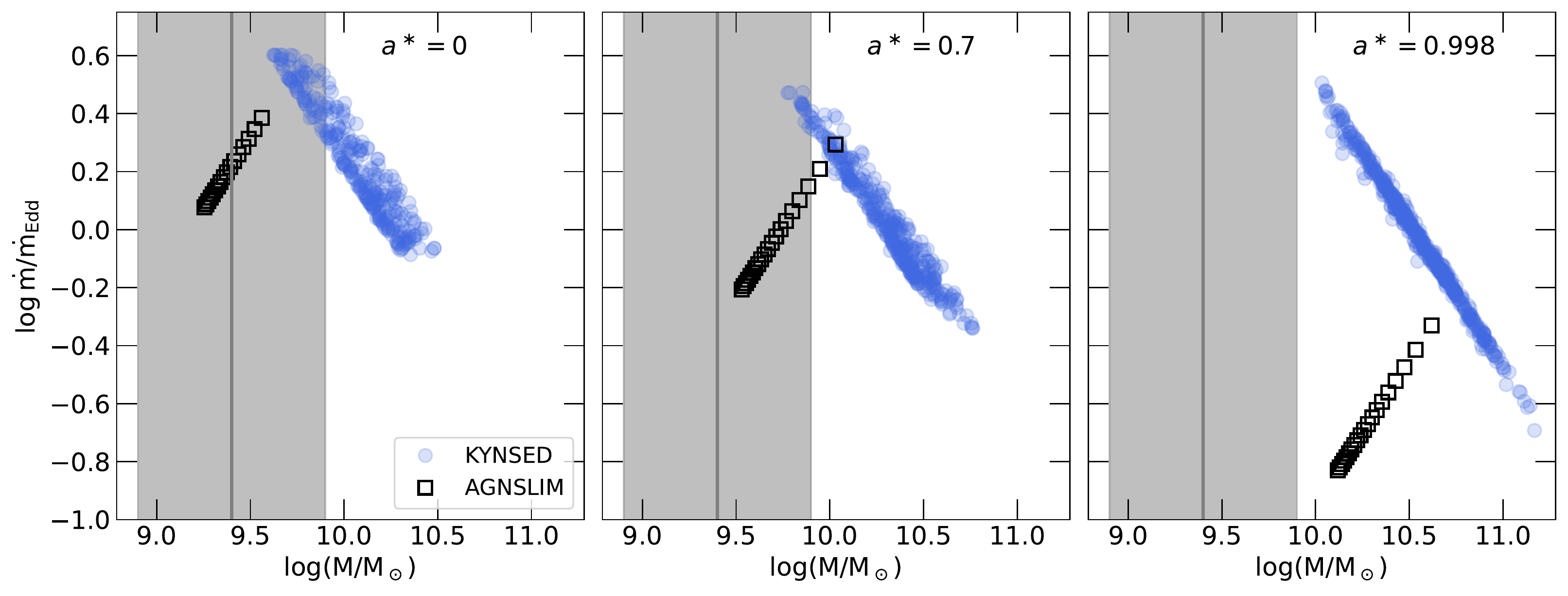}
\caption{Best-fit \mbh\ and \mdot\ obtained by fitting the SED of J1144 assuming $a^\ast =0, 0.7,$ and 0.998 (left to right panels) using {\tt KYNSED} (filled circles) and {\tt AGNSLIM} (open squares). The shaded grey area corresponds to the \mbh\ measurement reported by \citet{Onken22}, using the width of optical emission lines.}
\label{fig:massmdot}
\end{figure*}

Figure\,\ref{fig:massmdot} shows how the \mbh\ and \mdot\ change for different spins. Similarly to the {\tt KYNSED} model, the increase in spin results in a global increase \mbh\ and  decrease in \mdot. However, for a given spin value, both \mbh\ and \mdot\ decrease as $\cos \theta$ increases (see Fig.\,\ref{fig:cornerSED_agnslim}). This is due to the fact that, in {\tt AGNSLIM}, $\cos \theta$ acts as a normalisation factor of the flux, without any effect on the spectral shape. In this case, when $\cos \theta$ changes, both \mbh\ and \mdot\ adjust to compensate for the change in flux. The best-fit values of \mbh\ are consistent with the one derived by \cite{Onken22} for $a^\ast = 0$ and 0.7. A maximally spinning BH results in an \mbh\ larger by an order of magnitude compared to this value. The best-fit values of BH mass obtained using {\tt KYNSED} and {\tt AGNSLIM} are consistent within a factor  $\lesssim 2 $. However, the accretion rate changes significantly between the two models. This difference increases as the BH spin increases. Since the best-fit mass and accretion rate correlate with the color correction in {\tt KYNSED} (see Fig.\,\ref{fig:cornerSED}), only the results with $f_{\rm col} = 1$ should be considered to compare the results with {\tt AGNSLIM}. This corresponds to the lowest masses and highest accretion rates for a given spin in {\tt KYNSED}. In Appendix\,\ref{sec:comparison}, we compare the two models in detail for $\mdot = 0.1$ and 1. It is worth noting that, as discussed in \cite{Kubota19}, {\tt AGNSLIM} is consistent with a standard accretion disc for $\mdot \lesssim 2.39$ which is the case of J1144. {\tt AGNSLIM} predicts a larger UV emission than {\tt KYNSED}. The difference increases with spin. This is mainly due to the fact that {\tt AGNSLIM} does not take into consideration general relativity (GR) effects. Due to GR, a large amount of flux from the inner disc will end up in the BH, hence the difference between the two models. In order to compensate for this, fitting with {\tt AGNSLIM} results in lower mass and accretion rate compared to {\tt KYNSED}. For the non-spinning case, {\tt AGNSLIM} results in an accretion rate above the Eddington limit but below the critical limit of $\sim 2.39\, \dot{m}_{\rm Edd}$. For $a^\ast = 0.7$, \mdot\ gets below the Eddington limit for $\cos \theta > 0.4$. The maximally spinning case results in a sub-Eddington accretion rate for all cases. We note that this modelling also predicts a compact X-ray corona located within $10\,\rg\ $ of the BH for all spin values. The value of $R_{\rm h}$ decreases by increasing the spin.

We derive from this model a bolometric luminosity $L_{\rm bol} = 6.00_{-0.04}^{+0.11} \times 10^{47}\,\rm erg\,s^{-1}$, consistent with the value derived from the {\tt KYNSED} modelling. The measured values of $L_{\rm bol}$ and \mbh\ imply an Eddington ratio of $\lambda_{\rm Edd} = 2.08_{-0.78}^{+0.56}, 1.04_{-0.60}^{+0.36},$ and $0.27_{-0.16}^{+0.09}$, for $a^\ast = 0, 0.7$, and 0.998, respectively. If the BH spin were low/intermediate, this would imply a super-Eddington accretion rate consistent with the value derived  by \cite{Onken22}.

\section{Discussion}
\label{sec:discussion}

We have presented in this work the results from X-ray observations of J1144, the most luminous QSO in the last $\sim 9\,\rm Gyr$, using eROSITA, \swift, \xmm, and \nustar. Despite the fact that the source was not detected by ROSAT, it is currently detected by all of the observatories showing an X-ray variability by a factor of $\sim 10$ within a year. We also detected a shorter timescale variability of the order of $\sim 2.7$ within $\sim 40$\,days. The X-ray spectrum of this source can be well described using an absorbed power law with a high-energy cutoff. We also modelled the broadband SED of the source. Both a standard accretion disc irradiated by a point-like X-ray source, and a slim-disc emissivity profile could fit the observed SED equally well. This resulted in a bolometric luminosity of $\rm 6.2_{-0.5}^{+2.5} \times 10^{47}\,erg\,s^{-1}$ ($6.00_{-0.04}^{+0.11} \times 10^{47}\rm \,erg\,s^{-1}$) using {\tt KYNSED} ({\tt AGNSLIM}). This makes it the brightest QSO for $z\lesssim 1.3$, and among the most luminous 0.1\% known QSOs. 

\subsection{X-ray properties}
\label{sec:xrayprop}

\subsubsection{Coronal properties}

The X-ray spectrum of this source is consistent with a power law with a high-energy cutoff. Due to the quality of the data, we assumed that the photon index is constant for all the eRASS observations, and also constant during the recent monitoring in 2022. The photon indices from the two epochs are consistent within uncertainties suggesting a rather soft spectrum. Thanks to the \nustar\ observation, we are able to measure the high-energy cutoff. The measured value of $E_{\rm cut}$ depends on the employed model. For a simple absorbed power law we found $E_{\rm cut} = 68_{-23}^{+65}\, \rm keV$. However, when a reflection component is added this value reaches $23_{-6}^{+13}\,\rm keV$. This is due to the fact that the latter model assumes that part of the curvature in the hard X-rays is also due to the presence of the Compton hump (see left panel of Fig.\,\ref{fig:SEDmodel}), which shifts $E_{\rm cut}$ to a lower value. However, in both cases $E_{\rm cut}$ is one of the lowest measured in AGN \citep[see e.g.,][]{Kara17, Reeves21}. When placed in the $L_{\rm X}-E_{\rm cut}$ plane, the source falls well in the limited region allowed to avoid runaway pair production, at its corresponding X-ray luminosity \citep[e.g.,][]{Fab15, Fabian17,Lanzuisi19}. This may suggest that the corona of J1144 is a pair dominated hybrid plasma \citep[see][]{Fabian17}.

\begin{figure}
\centering
\includegraphics[width=0.95\linewidth]{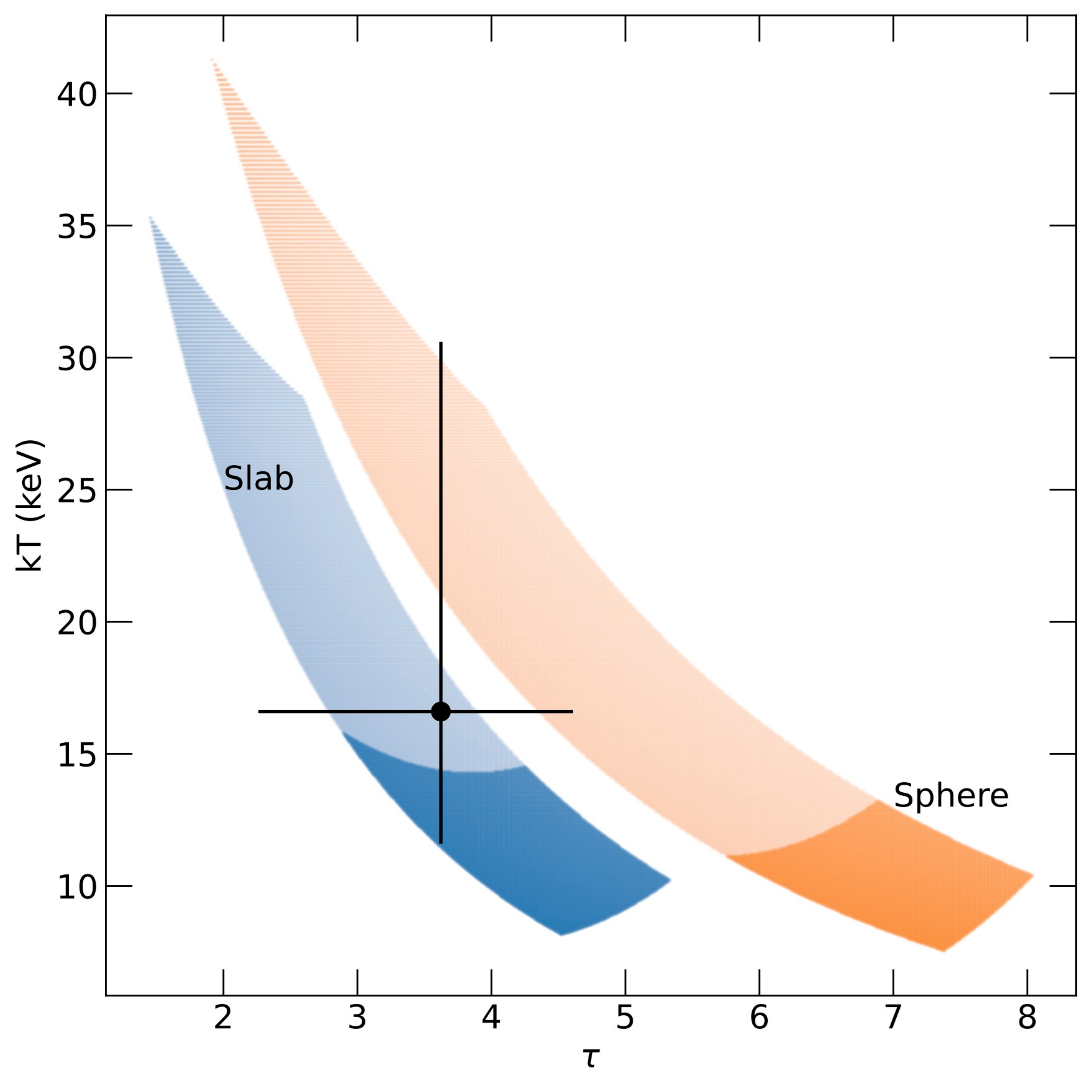}

\caption{Electron temperature ($kT$) vs optical depth ($\tau$) of the corona obtained by assuming Comptonization and mapping the $\Gamma-E_{\rm cut}$ plane into this plane, following \citet{Middei19}. We assumed a slab and spherical geometry (blue and orange, respectively). The darker regions correspond to the limits if $E_{\rm cut}$ is obtained from the {\tt KYNSED} model. The lighter regions are obtained by extending the upper limit of $E_{\rm cut}$ to $70\,\rm keV$, as a conservative limit, to match the results obtained from fitting a simple power law with a high-energy cutoff model. The black circle correspond to the estimates of $kT$ and $\tau$ obtained from fitting the SED with {\tt AGNSLIM}, which assumes a slab geometry.}
\label{fig:taukt}
\end{figure}

\begin{figure}
\centering
\includegraphics[width=0.98\linewidth]{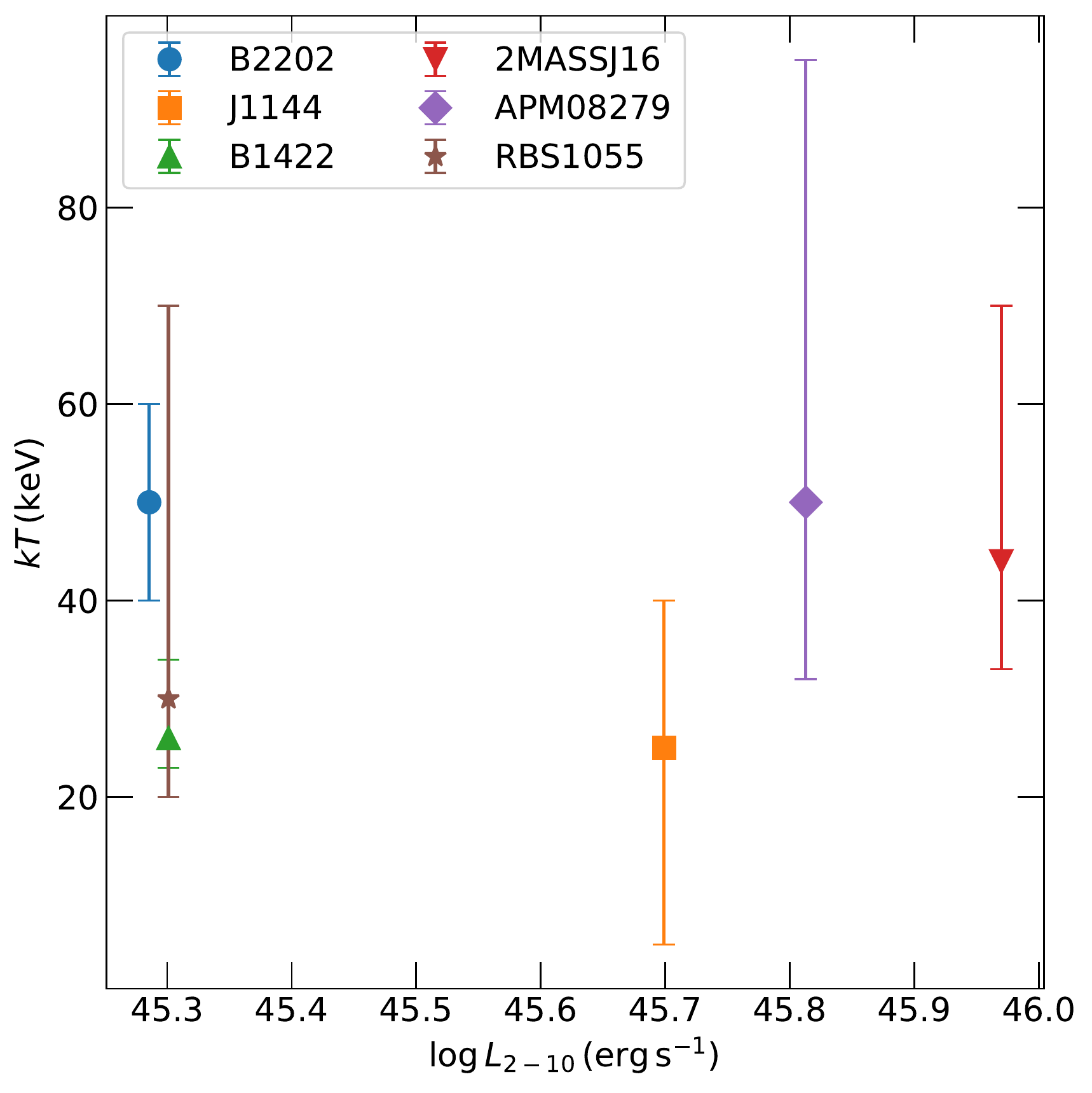}

\caption{Comparison of J1144 to other quasars from the literature, with $L_{2-10} > 10^{45}\, \rm erg\,s^{-1}$, in the $kT-L_{2-10}$ plane.}
\label{fig:ktLx}
\end{figure}

Using Eqs.\,($2-5$)\footnote{We note that a minus sign is missing from Eq.\,(5) in \cite{Middei19}. The correct expression is $\beta(\tau) = -3.35 + 1.3 \tau - 0.11  \tau^2$.} from \cite{Middei19}, we mapped our results from the $\Gamma-E_{\rm cut}$ plane in the $kT-\tau$ plane. We used the results from modelling the broadband SED, with $\Gamma$ in the range $2-2.4$ and $E_{\rm cut}$ in the range $15-70$\,keV, assuming a slab and a spherical geometry of the corona. We consider a conservative upper limit on $E_{\rm cut}$ to take into account the value measured using a simple power-law model. The results are shown in Fig.\,\ref{fig:taukt}. For a slab geometry, we find $\tau$ in the range $1.5-5.4$ and $kT$ in the range $5-35\,\rm keV$. For a spherical geometry, we find $\tau$ in the range $2-8$ and $kT$ in the range $7-40\,\rm keV$. We also used the best-fit temperature and photon index derived from {\tt AGNSLIM} ($kT =16_{-5}^{+14}\,\rm keV, \Gamma = 2.09 \pm 0.04$), which assumes a slab geometry, to derive the optical depth. Using Eq. (2) of  \cite{Middei19}, we obtain $\tau = 3.6_{-1.4}^{+1.0}$ (we considered the uncertainty on $kT$ only). These values of $kT$ and $\tau$ are in agreement with the ones derived from {\tt KYNSED} (see Fig.\,\ref{fig:taukt}). Our results add another QSO to the highly accreting sources with a low coronal temperature \citep[e.g., Ark\,564 and PDS 456][]{Kara17,Reeves21}. More recently, \cite{Tortosa2023} also found low temperatures in two rapidly accreting AGN  Mrk\,382, IRAS\,04416+1215. In Fig.\,\ref{fig:ktLx}, we compare J1144 to other sources with $L_{\rm 2-10} > 10^{45}\,\rm erg\,s^{-1}$, namely: B2202--209 \citep{Kammoun17}, 2MASS\,J1614346+470420  and B1422+231 \citep{Lanzuisi19}, APM\,08279+5255 \citep{Bertola22}, and RBS\,1055 \citep{Marinucci2022}. The coronal temperature in J1144 is broadly consistent with these sources, being among the lowest. We note also that, as mentioned earlier, the two models used in this work provide an estimate of the size of the X-ray corona. The measured sizes are listed in Tables\,\ref{tab:SED}-\ref{tab:SED_agnslim}. Both models predict a compact corona with radius smaller than $\sim 10\,r_{\rm g}$ ({\tt KYNSED} gives an upper limit of $25\,r_{\rm g}$).

\begin{figure}
\centering
\includegraphics[width=0.98\linewidth]{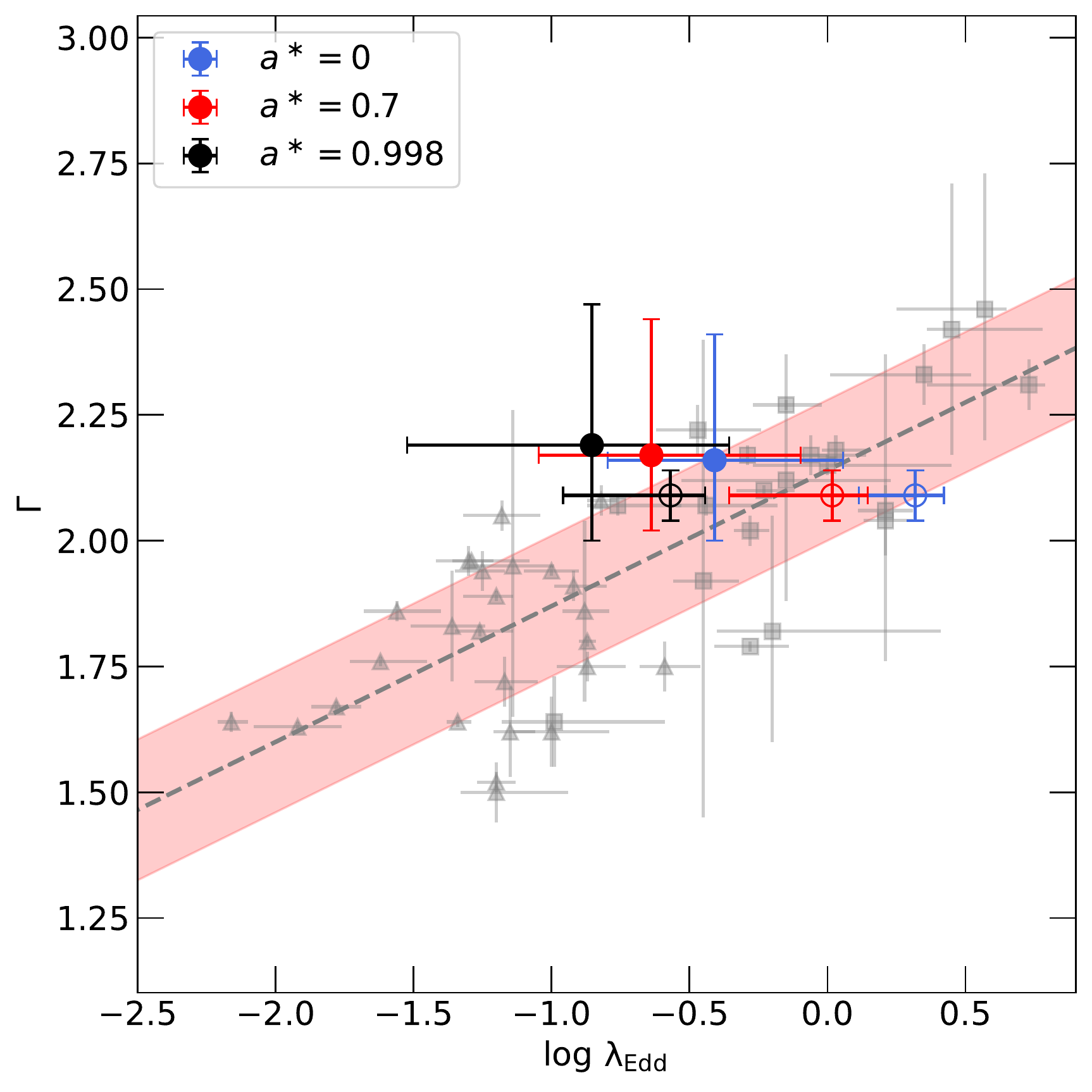}

\caption{Photon index versus Eddington ratio assuming $a^\ast = 0, 0.7,$ and 0.998 (blue, red, and black, respectively). Filled and open circles correspond to the results obtained using {\tt KYNSED} and {\tt AGNSLIM}, respectively. Triangles and squares correspond to the data from \citet{Liu21} for sub- and super-Eddington sources, respectively. The dashed line and the shaded area represent the best-fit to the \citet{Liu21} data and the corresponding $1\sigma$ scatter, respectively.}
\label{fig:GammaLedd}
\end{figure}

Various works have presented a positive correlation between the photon index and the Eddington ratio in QSOs \citep[e.g.,][]{Shemmer08,Risaliti09,Brightman13,Liu21}. This correlation is usually explained by the fact that at high Eddington ratios the UV/optical emission from the accretion disc is enhanced which leads to a more efficient Compton cooling of the corona, decreasing $kT$, which leads to a softening of the X-ray spectrum (an increase in the photon index). Figure\,\ref{fig:GammaLedd} shows $\Gamma$ versus $\log \lambda_{\rm Edd}$ for the different spin values considered in this work using {\tt KYNSED} and {\tt AGNSLIM}. We compare these results to the data obtained by \cite{Liu21}. J1144 is consistent with the observed $\Gamma - \log \lambda_{\rm Edd}$ correlation, for both models and all spin values.

\subsubsection{X-ray reflection}

No strong signature of the presence of reflection can be inferred from the current spectra. However, as mentioned in Appendix\,\ref{sec:linescan}, including an emission line at $\sim 6.5\,\rm keV$ improves the fit by $\Delta \chi^2 \simeq -9$. The equivalent width of the line $(\rm EW = 117 \pm 44\,\rm eV)$ is in agreement with the expected one, of 150\,eV, for solar abundance and a $2\pi$ covering \citep{Geo91}. We note that this value is larger than the one expected from the Iwasawa-Taniguchi effect \citep[see e.g.,][]{Iwasawa93,Bianchi07} for the luminosity of this source. Similar large equivalent widths have also been seen in a few powerful QSOs \citep[e.g.,][]{Krumpe10,Marinucci2022}.

{\tt KYNSED} includes a disc reflection component with a self-consistently calculated ionisation profile. The model provides the ionisation parameter ($\xi_{\rm in}$) at the inner edge of the disc as an output. These values are listed in Table\,\ref{tab:SED}. The model suggests a low ionisation state of the disc, consistent with neutral, for all of the spin values. In order to estimate the importance of reflection in the model, we calculate the ratio of the reflection component to the power law component for each of the fits in the $1.6-16\,\rm keV$ observed range (equivalent to $3-30\,\rm keV$, rest frame). Then, to compare this ratio to the commonly used reflection fraction ($\mathcal{R}$) we used the relation between this flux ratio and $\mathcal{R}$ derived in Fig.\,9 of \cite{Kammoun2020}, for neutral reflection:

\begin{equation}
    \log \mathcal{R} = (1.38 \pm 0.18) \log \left( \frac{F_{\rm ref}}{F_{\rm PL}}\right)_{3-30} + (0.65 \pm 0.06).
\end{equation}

\noindent This results in $\mathcal{R} = 0.82_{-0.27}^{+0.66} $ considering all the spin values. The values of $\mathcal{R}$ for each spin are shown in Table\,\ref{tab:SED}. We tested the reflection spectrum by modelling the \xmm\ and \nustar\ spectra only using {\tt Relxill} \citep{Dauser13, Dauser16}. This gives consistent results with the modelling using {\tt KYNSED}. The best fit photon index, cutoff energy, and reflection fraction are $\Gamma = 2.17 \pm 0.13$, $E_{\rm cut} = 48_{-11}^{+33}\,\rm keV$, and $\mathcal{R} = 1.2_{-0.8}^{+1.1}$. Deeper exposures are required to confirm the presence of the reflection with a higher confidence.

It is worth noting that no soft X-ray excess has been seen in this source. In particular, the spectrum in eRASS1, when the source was at its highest flux with low intrinsic absorption, is consistent with a simple power law. This, interestingly, rules out the presence of any strong soft component in J1144.

\begin{figure*}
\centering
\includegraphics[width=0.95\linewidth]{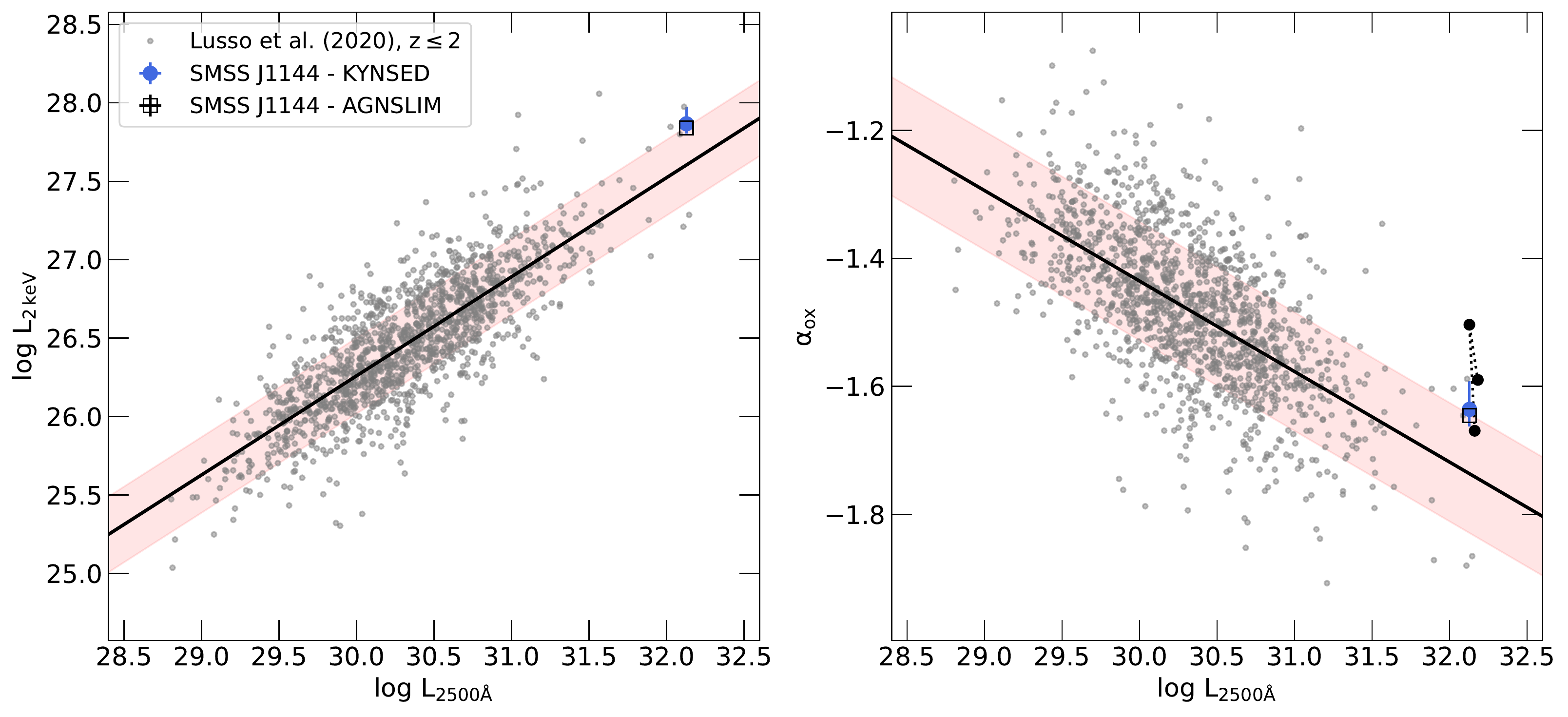}

\caption{$L_{\rm 2 keV}$ vs $L_{2500\AA}$ (left) and $\alpha_{\rm ox}$ vs $L_{2500\AA}$ (right) for the QSOs at $z \leq 2$ presented in \citet{Lusso2020} (grey dots). The solid line represents the best-fit straight line to these data. The red shaded area represents the $1\sigma$ scatter around the best-fit model. The estimates of J1144 from the SED fitting are shown as a blue circle for {\tt KYNSED} and as a empty black square for {\tt AGNSLIM}. The black connected circles show the variability of $\alpha_{\rm ox}$ during the \swift\ monitoring (see Sec.\,\ref{sec:aox} for details).}
\label{fig:LxLuv}
\end{figure*}

\subsubsection{X-ray to UV/optical ratio}
\label{sec:aox}

\begin{table*}
\centering
\caption{Bolometric luminosity, unabsorbed $2-10$\,keV luminosity, the optical-to-X-ray ratio ($\alpha_{\rm ox}$), and the Eddington ratio obtained by modelling the SED of J1144. We show the values obtained assuming different spin values. We also show the results for all spins together. Similar to Table\,\ref{tab:SED}, the uncertainties represent the lowest/highest values obtained from the SED modelling.}
\label{tab:SEDnew}
\begin{tabular}{lcccccccc} 
\hline \hline			
& \multicolumn{4}{c}{{\tt KYNSED }} & \multicolumn{4}{c}{{\tt AGNSLIM}} \\[0.1cm]
	&		$a^\ast = 0$					&			$a^\ast = 0.7$				&			$a^\ast = 0.998$	& All			&		$a^\ast = 0$					&			$a^\ast = 0.7$				&			$a^\ast = 0.998$	& All\\ \hline
$L_{\rm bol}\,\rm(10^{47}\,erg\,s^{-1})$	&	$6.19_{-0.33}^{+0.58}$	&	$6.20_{-0.34}^{+0.81}$	&	$6.26_{-0.45}^{+1.37}$	&	$6.21_{-0.38}^{+1.42}$	&	$5.97_{-0.01}^{+0.01}$	&	$6.00_{-0.01}^{+0.01}$	&	$6.10_{-0.01}^{+0.01}$	&	$6.00_{-0.04}^{+0.11}$	\\[0.1cm]
$L_{\rm 2-10}\,\rm(10^{45}\,erg\,s^{-1})$	&	$5.11_{-0.29}^{+0.47}$	&	$5.09_{-0.28}^{+0.49}$	&	$5.08_{-0.31}^{+0.52}$	&	$5.09_{-0.32}^{+0.51}$	&	$4.99_{-0.01}^{+0.01}$	&	$4.99_{-0.01}^{+0.01}$	&	$4.99_{-0.01}^{+0.01}$	&	$4.99_{-0.01}^{+0.01}$	\\[0.1cm]
$\alpha_{\rm ox}$	&	$-1.63_{-0.02}^{+0.04}$	&	$-1.63_{-0.02}^{+0.04}$	&	$-1.64_{-0.03}^{+0.04}$	&	$-1.63_{-0.03}^{+0.04}$	&	$-1.65_{-0.01}^{+0.01}$	&	$-1.65_{-0.01}^{+0.01}$	&	$-1.65_{-0.01}^{+0.01}$	&	$-1.65_{-0.01}^{+0.01}$	\\[0.1cm]
$\lambda_{\rm Edd}$	&	$0.39_{-0.23}^{+0.75}$	&	$0.23_{-0.14}^{+0.57}$	&	$0.14_{-0.11}^{+0.30}$	&	$0.22_{-0.19}^{+0.92}$	&	$2.08_{-0.78}^{+0.56}$	&	$1.04_{-0.60}^{+0.36}$	&	$0.27_{-0.16}^{+0.09}$	&	$0.97_{-0.85}^{+1.66}$	\\ \hline
																	
\end{tabular}
\end{table*}

Comparing the X-ray to the bolometric luminosity inferred from the broadband SED modelling we find a $L_{\rm bol}/L_{\rm 2-10} \sim 120$. Such a large value is typically seen in bright QSO \citep[e.g.,][]{Lusso2012,Duras20}. We measured the unabsorbed specific luminosity at $2500\,\AA$ and $2\,\rm keV$, using the best-fit SED models. For comparison, we plot, in the left panel of Fig.\,\ref{fig:LxLuv}, $\log L_{\rm 2keV}$ vs $\log L_{\rm 2500\AA}$ from \cite{Lusso2020} in grey. We fitted these data with a straight line using the ordinary least square method \citep[OLS(Y|X);][]{Isobe1990}. The black solid line corresponds to the best-fit line. J1144 agrees well with this relation within $1\sigma$ (red shaded area). We also estimated the X-ray to UV ratio, defined as $\alpha_{\rm ox} = 0.3838 \log \left(  L_{\rm 2keV}/L_{\rm 2500\AA} \right)$, to be $-1.64_{-0.04}^{+0.06}$ ($-1.65\pm 0.01$) using {\tt KYNSED} ({\tt AGNSLIM}). The right panel of Fig.\,\ref{fig:LxLuv} shows $\alpha_{\rm ox}$ vs $\log L_{\rm 2500\AA}$, also from \cite{Lusso2020}. We also fit a straight line to the relation, and we find that J1144 agrees with it within $1\sigma$. Considering the intrinsic absorption will result in $\alpha_{\rm ox} = -1.7$, still consistent with the \cite{Lusso2020} results. In addition, we estimated $\alpha_{\rm ox}$ during the \swift\ monitoring. We used the UVOT/M2 filter to scale $L_{\rm 2500\AA}$ and considered the observations where XRT and UVOT/M2 data are available (XRT/O1, O9, and O10). The estimated values of $\alpha_{\rm ox}$ are shown as black connected circles in Fig.\,\ref{fig:LxLuv}. Interestingly, in all these observations, the source never reaches an X-ray weak state\footnote{X-ray weakness is usually estimated using $\Delta \alpha_{\rm ox}$ which is the difference between the estimated and the predicted $\alpha_{\rm ox}$ for a given luminosity. Sources with $\Delta \alpha_{\rm ox} \leq -0.3$ can be reasonably classified as X-ray weak.}, contrary to other variable high-Eddington sources \citep[see e.g.,][]{Laurenti21}.

\subsection{Outflow signature}
\label{sec:outflow}

\begin{figure}
\centering
\includegraphics[width=0.98\linewidth]{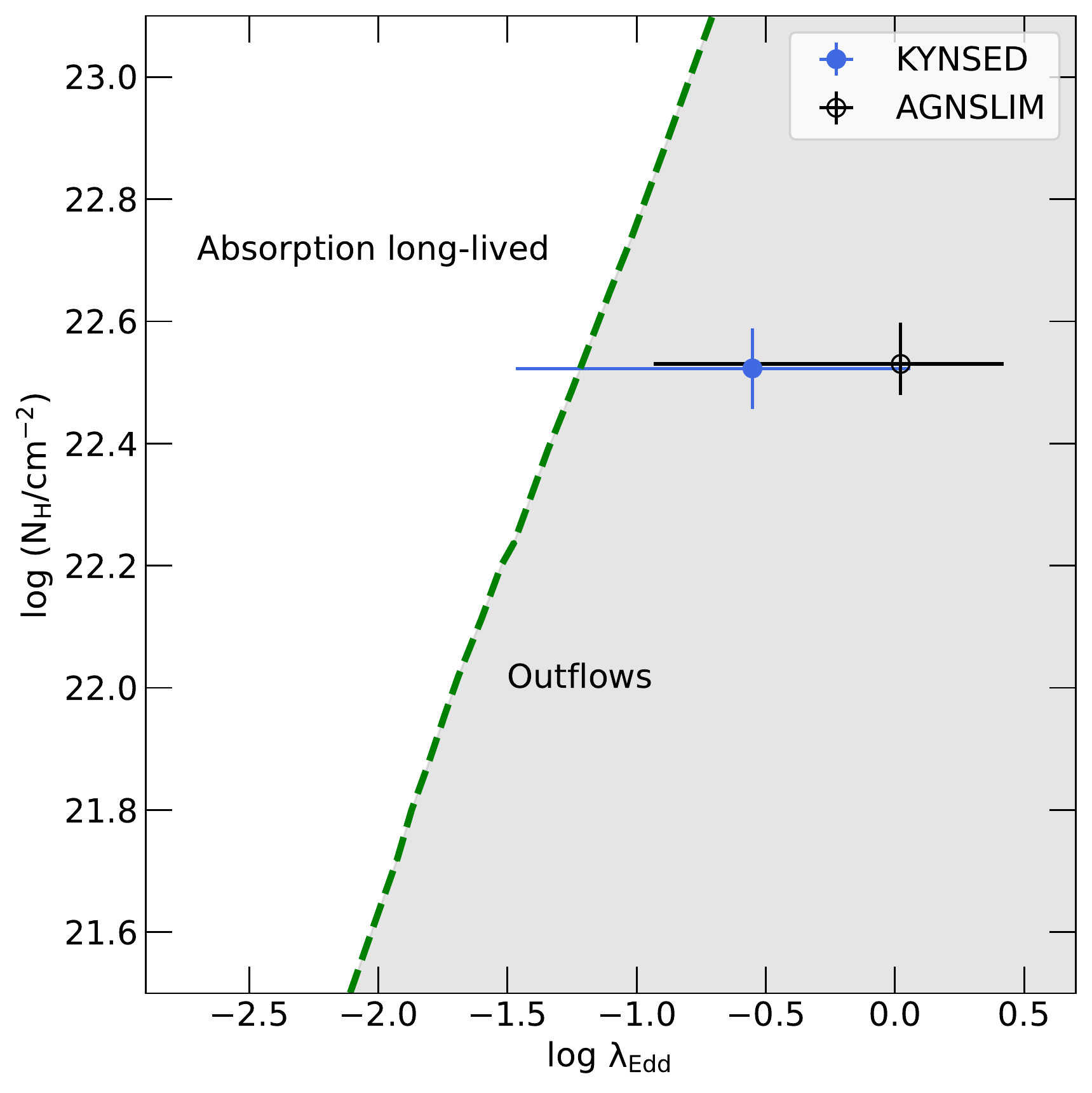}

\caption{Intrinsic absorption column density versus the Eddington ratio obtained from modelling the SED of J1144 using {\tt KYNSED} (blue circle) and {\tt AGNSLIM} (black empty square). The green dashed line shows the effective Eddington limit above which dusty clouds \citep[with standard ISM grain abundance; adapted from][]{Fabian2009} see the AGN as being effectively above the Eddington limit. Long-lived absorbing clouds can only occur for $N_{\rm H}$ above this line. J1144 falls within the outflow region shown in grey.}
\label{fig:nhledd}
\end{figure}

The X-ray spectra of the source do not show any evidence of absorption by winds or ultra fast outflows. The absorption line detected at $\sim 1.3\,\rm keV$ could be due to some intrinsic absorbing material. However, identifying the origin of this line is hard. Similar features can also be seen in the eROSITA (eRASS2 and eRASS5) and \swift/XRT spectra. The quality of the data does not allow us to confirm the existence of these features with high confidence. However, the ``transient-like'' aspect of these features could hint at an outflowing origin.

\cite{Fabian2006, Fabian2008, Fabian2009} discussed that, in the presence of dust, the gas couples with the dust grains via Coulomb interaction and the cross-section for the interaction with photons is enhanced. Thus, the effective Eddington limit, for which the outward radiation pressure on gas exceeds the inward gravitational pull, is much lower for dusty gas than for ionized dust-free gas \citep[e.g.,][]{Laor93,Scoville95}. This implies that AGN considered as sub-Eddington using the standard definition may nevertheless exceed the effective Eddington limit for substantial column densities of dusty gas. In this case, long-lived, stable clouds can survive radiation pressure only in a regime lower than the effective Eddington limit. Otherwise, the gas seeing the nucleus above the effective Eddington limit is expelled, and should be experiencing outflows. In this case absorption may be transient or variable. \cite{Fabian2008, Fabian2009} used {\tt CLOUDY} to derive a limit between the long-lived and the outflow absorption in the $N_{\rm H}-\lambda_{\rm Edd}$ plane. This limit is shown in Fig.\,\ref{fig:nhledd}. J1144 resides on the right-hand side of that limit, suggesting that the absorption in this source is due to outflow. The fact, that the absorption measured from the eROSITA spectra is much lower than the one measured during the more recent monitoring by more than an order of magnitude supports this hypothesis, as this absorption is expected to be variable. Furthermore, our modelling of the SED requires a partial covering of the X-ray source without any additional absorption for the UV/optical. This could indicate that this absorption is located closer to the BH, and is of an outflow origin. \cite{Baskin2018} estimated the inner disc radius ($R_{\rm in} = 0.018  L_{46}^{0.5}\,\rm pc$, where $L_{46}$ is the bolometric luminosity in units of $10^{46}\,\rm erg\,s^{-1}$) which can have a dusty atmosphere, so dusty wind could be launched at a distant between $R_{\rm in}$ and the sublimation radius. This results in $R_{\rm in} = 0.04\,\rm pc$. Considering $\log M_{\rm BH}/M_\odot$ between 9.5 and 11, we estimate $R_{\rm in}$ to be between $\sim 30-900\,\rg$. Supporting further the possibility of the absorption being connected to disc winds. Better quality data are needed to confirm this. High quality X-ray and UV/optical spectra will be crucial to detect the possible presence of outflows in this source, and to study the variability of absorption. In particular, the next generation of X-ray microcalorimeter like \textit{XRISM}/Resolve \citep{Tashiro18} and \textit{Athena}/X-IFU \citep{Barret2023} will unveil more secrets about this and other sources at comparable redshift. This will help us to better understand the evolution of such massive and rapidly accreting black holes.

\subsection{Black hole mass}
\label{sec:bhmass}

The inferred BH mass depends strongly on the assumed SED model. An X-ray illuminated standard accretion disc predicts higher mass values compared to a model assuming a slim disc emissivity profile, which gives a mass estimate closer to the single-epoch value obtained by measuring line width in \cite{Onken22}. In fact, the single-epoch mass estimates are subject to various biases and uncertainties \citep[e.g.,][]{Shen2013}. In particular, the virialized masses estimated are thought to be underestimated in sources like J1144, where radiation pressure is important \citep[][]{Marconi08, Marconi09}. It is also worth noting that the relations used by \cite{Onken22} represent an extrapolation by almost an order of magnitude in luminosity compared to the $\rm H\beta$ reverberation mapping sample of \cite{Bentz13}. Thus, it may be plausible that the true mass of the source is larger than the value estimated by \cite{Onken22}.

Both models used in this work predict a relatively large mass for a spinning BH. It is also worth noting that, although {\tt AGNSLIM} adopts a slim disc emissivity profile, it has several assumptions that could affect the measured \mbh. For instance, the model neglects all the general relativity effects which could alter emission from the innermost regions of the system, especially for sources where the corona is quite compact similar to J1144. In addition, {\tt AGNSLIM} does not account properly for the inclination of the system that could affect the spectral shape \citep[not only the overall flux; see Fig. A6 in][]{Dovciak22}. It also neglects the presence of any reprocessing of the X-rays by the disc, which could affect the X-ray spectrum as well as the disc emission by heating its surface \citep[see][]{Kammoun21theory, Dovciak22}. As for {\tt KYNSED}, the BH mass inferred from this model is highly affected by the assumed value of \fcol. However, the true value of \fcol\ is quite uncertain, and depends itself on the BH mass and accretion rate \citep[see e.g.,][]{Davis19}.

The current data do not allow us to distinguish between the two models. However, the source is confirmed to be in the high-mass regime with $\log \mbh/M_\odot \gtrsim 9.5$. \cite{King2016} estimated the maximum physical limit of mass that an SMBH can reach through luminous accretion of gas as a function of the BH spin (see their Figure\,1). All of the estimated values of the BH mass in J1144 (using both models) lie below this limit. We note that, in all cases, the main conclusions of our work will not be strongly affected by the exact value of \mbh.

\subsection{Variability}
\label{sec:variability}
As mentioned earlier, the ROSAT upper limit is consistent with the low-flux state seen in the data presented in this paper. Moreover, we used HILIGT to derive upper limits for the \xmm\ Slew catalogue. These limits are not very constraining as they are consistent with the high-flux state of the source. Thus, we cannot conclude on the X-ray variability of the source over a timescale of a couple of decades. In other terms, we cannot confirm whether the ROSAT non-detection is due to an intrinsic X-ray weakness of the source or due to the flux limit of the observations.

On the shorter timescales, in addition to the variability seen in absorption, the source also exhibits intrinsic flux changes on timescales of days to years that could reach a factor of $\sim 10$ as seen in the eROSITA data. The shortest max-to-min change seen during the monitoring of the source is of the order of 2.7 over $\sim 17$\,days (observed). Based on the {\tt KYNSED} results, this would correspond to $\sim 1.6-24.6$ times the light-crossing time per gravitational radius ($t_{\rm cross} = G\mbh/c^3$), in the rest frame of the source after correcting for time dilation due to the cosmological redshift. This would increase up to $\sim 66\,t_{\rm cross}$ for $\log \mbh/M_\odot = 9.4$. The range is quite uncertain due to the uncertainty on the mass. However, in all cases, this variability timescale is longer than the light crossing time. Thus, the X-ray variability could simply originate from intrinsic changes in the luminosity of the X-ray corona. A lower amplitude variability is also seen in the in the UV/optical range. While this is very unlikely due to intrinsic changes in the accretion disc, the observed variability may be well driven by thermal reverberation as the disc responds to the X-ray variability \citep[e.g.,][]{Kammoun21theory}. It is worth noting that the quality of the data does not allow us to constrain any spectral changes that could occur in the source. Addressing all these points requires a more intense monitoring campaign and deeper observations in X-ray/UV/optical.

\section{Conclusion}

In this paper we analysed the X-ray spectra of J1144 from five eROSITA observations performed between the end of 2019 and the end of 2021. In addition, we analyse the results obtained from a recent monitoring of the source using \swift, \xmm, and \nustar. The source shows a large X-ray variability that is due to intrinsic changes in the X-ray luminosity of the source accompanied with changes in the absorption in the line of sight. This absorption could be due to a radiatively driven outflow material. The observed SED of J1144 could be fitted equally with a standard accretion disc around a BH with a mass of a few times $10^{10}\,M_\odot$, and with a slim disc model assuming a smaller BH mass of the order of a few times $10^{9}\,M_\odot$. In both cases, the source seems to accrete at a rate larger than 40\% of the Eddington limit. If we assume a low BH spin the accretion rate can even exceed the Eddington limit. Assuming a Comptonisation model, we measure the coronal electron temperature to be of the order of $\sim 10-40$\,keV. With a bolometric luminosity of $6.2 \times 10^{47}\,\rm erg\,s^{-1} $, this source is the most luminous QSO in the last 9\,Gyr. Interestingly, the optical-X-ray properties of the source are different than many high-Eddington sources. Notably, the measured $\alpha_{\rm ox}$ value is consistent with standard radio-quiet QSOs rather than high-Eddington QSOs which tend to be X-ray weak. This could hint towards a sub-Eddington accretion rate, thus a non-zero BH spin and a large mass. Moreover, the source shows a hint of an Fe K$\alpha$ line with an equivalent width of $117 \pm 44\,\rm eV$, that is larger than what is expected for sources with a similar X-ray luminosity. Modelling the SED by including an X-ray reflection results in a reflection fraction of the order of unity.

Further deeper X-ray and UV/optical observations are needed to measure more accurately the nature of the absorption in this source and its variability. In addition, this will help confirm the presence of absorption features in the soft X-rays, check for further signatures of outflow, and better understand the origin of the large variability seen in this source. These observations will give us a glimpse at what happens in very luminous QSOs at cosmic noon, while requiring moderate observing time. This will allow us to better understand the growth of such massive black holes and study the connection between the activity of the central engine and its environment.

\section*{Acknowledgements}

We thank the PI of \nustar, Fiona A. Harrison, the Project Scientist of \xmm, Norbert Schartel, and the PI of \swift, Brad Cenko, as well as the planning staff of the three observatories for making these observations possible. 

This research is based on observations obtained with \xmm, an ESA science mission with instruments and contributions directly funded by ESA Member States and NASA. This research made use of data from the \nustar\ mission, a project led by the California Institute of Technology, managed by the Jet Propulsion Laboratory, and funded by NASA. This research made use of Astropy, a community-developed core Python package for Astronomy \citep{2018AJ....156..123A, 2013A&A...558A..33A}. This research made use of SciPy \citep{Virtanen_2020}. This research made use of NumPy \citep{harris2020array}. This research made use of matplotlib, a Python library for publication quality graphics \citep{Hunter:2007}. This research made use of XSPEC \citep{Arnaud96}. We acknowledge the use of public data from the \swift\ data archive.

This work is based on data from eROSITA, the soft X-ray instrument aboard SRG, a joint Russian-German science mission supported by the Russian Space Agency (Roskosmos), in the interests of the Russian Academy of Sciences represented by its Space Research Institute (IKI), and the Deutsches Zentrum für Luft- und Raumfahrt (DLR). The SRG spacecraft was built by Lavochkin Association (NPOL) and its subcontractors, and is operated by NPOL with support from the Max Planck Institute for Extraterrestrial Physics (MPE). The development and construction of the eROSITA X-ray instrument was led by MPE, with contributions from the Dr. Karl Remeis Observatory Bamberg \& ECAP (FAU Erlangen-Nuernberg), the University of Hamburg Observatory, the Leibniz Institute for Astrophysics Potsdam (AIP), and the Institute for Astronomy and Astrophysics of the University of Tübingen, with the support of DLR and the Max Planck Society. The Argelander Institute for Astronomy of the University of Bonn and the Ludwig Maximilians Universität Munich also participated in the science preparation for eROSITA. The eROSITA data shown here were processed using the eSASS/NRTA software system developed by the German eROSITA consortium.

ESK, DB, and POP acknowledge financial support from the Centre National d’Etudes Spatiales (CNES). ZI acknowledges funding from the ORIGINS cluster funded by the Deutsche Forschungsgemeinschaft (DFG, German Research Foundation) under Germany's Excellence Strategy – EXC-2094 – 390783311. POP and SB acknowledge financial support from the French National High Energy Program (PNHE) of CNRS. EP acknowledges financial support from PRIN MIUR project ``Black Hole winds and the Baryon Life Cycle of Galaxies: the stone-guest at the galaxy evolution supper'', contract \#2017PH3WAT. 

We thank the referee, Chris Done, for her useful comments and discussions about {\tt AGNSLIM}. We thank Riccardo Middei for clarifications about the results from the MoCA code. 

\section*{Data Availability}
The \swift, \xmm, and \nustar\ data are all available on the archives corresponding to each of the observatories. 
Data from eRASS1 are planned to be publicly released in Spring 2023, with the rest of the eROSITA data used in this paper (eRASS2--5) being made available two years later. 



\bibliographystyle{mnras}
\bibliography{reference} 


\appendix




\section{Line scan}
\label{sec:linescan}

\begin{figure}
\centering
\includegraphics[width=1\linewidth]{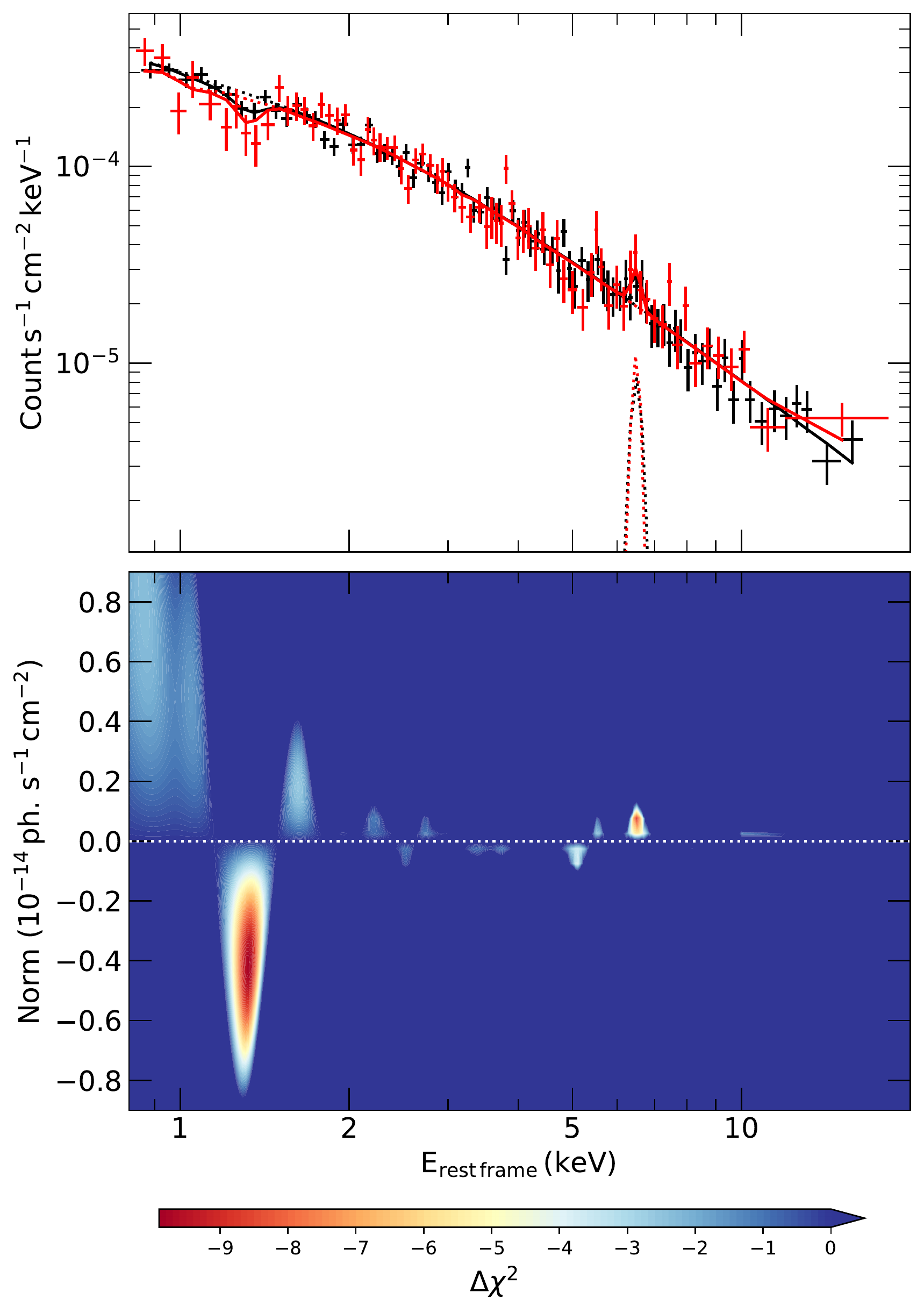}
\caption{Top: XMM-Newton spectra (PN/MOS in black/red)  fitted with a power law model, adding two Gaussian lines at 1.3\,keV (in absorption) and 6.5\,keV (in emsission), rest frame. Bottom: the improvement of the fit of the \xmm\ spectra, obtained by adding an absorption/emission line to the model.}
\label{fig:linescan}
\end{figure}

Despite the fact that an absorbed power law model explains the data well, some residuals in absorption (emission) could be seen at $\sim 1.3$\,keV ($6.5$\,keV), rest frame, in the \xmm\ spectra. We performed a scan of the \xmm\ data in the full observed range to check for the significance of the lines by adding a Gaussian line in emission or absorption with a fixed width $\sigma = 0.05\,\rm keV$. The results, presented in Fig.\,\ref{fig:linescan}, show an improvement of the quality of the fit by $\Delta \chi^2 = -9.9$ and $-8.9$ (for two additional free parameters) by adding an absorption line at $1.32$\,keV and emission at $6.5$\,keV. Refitting the \xmm\ spectra by adding these two lines and letting the width as a free parameter results in equivalent widths $EW_{\rm abs}(\rm 1.3\,keV) = -25\pm 10\,\rm eV$ and $EW_{\rm em}(\rm 6.5\,keV) = 117 \pm 44\,\rm eV$, with a total improvement of the fit by $\Delta \chi^2= - 19.5$ ($\sim 3\sigma$). The fit obtained by adding these features is shown in the upper panel of Fig.\,\ref{fig:linescan}. It is worth noting that the origin of the absorption line may not be identified. However, the emission line at 6.5\,keV could be associated with the Fe K$\alpha$ emission line.

\section{Residuals}

We show in this appendix the residuals obtained by modelling the X-ray spectra of J1144 from \swift, \xmm, and \nustar\ observations as shown in Section\,\ref{sec:monitoring}.

\begin{figure*}
\centering
\includegraphics[width=0.95\linewidth]{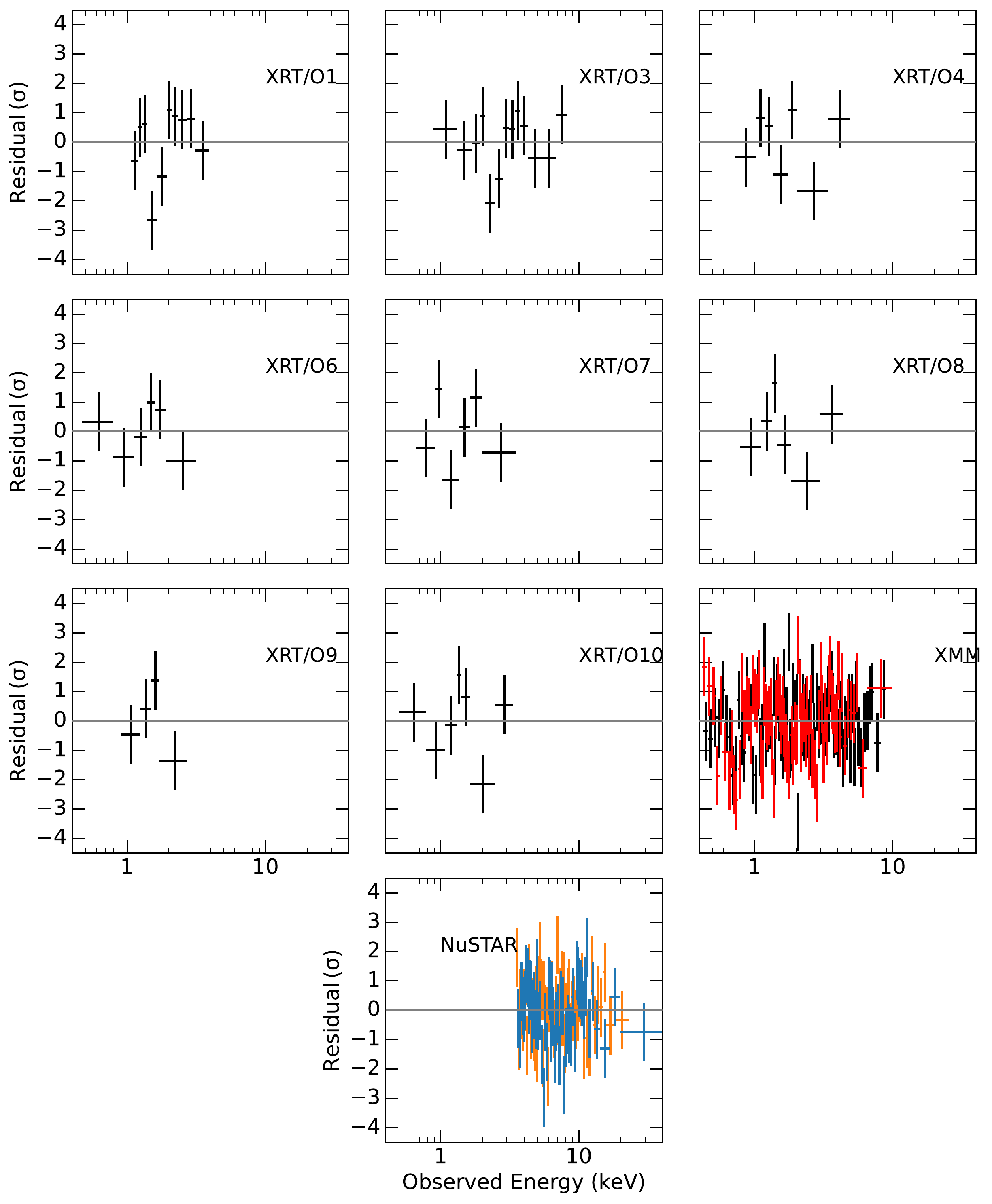}
\caption{Residuals obtained by modelling the X-ray spectra of J1144 obtained from \swift, \xmm, and \nustar\ observations using an aborbed power law with a high-energy cutoff (see Sec.\,\ref{sec:monitoring}).}
\label{fig:xrayresid}
\end{figure*}

\section{Contours}

\begin{figure*}
\centering
\includegraphics[width=0.95\linewidth]{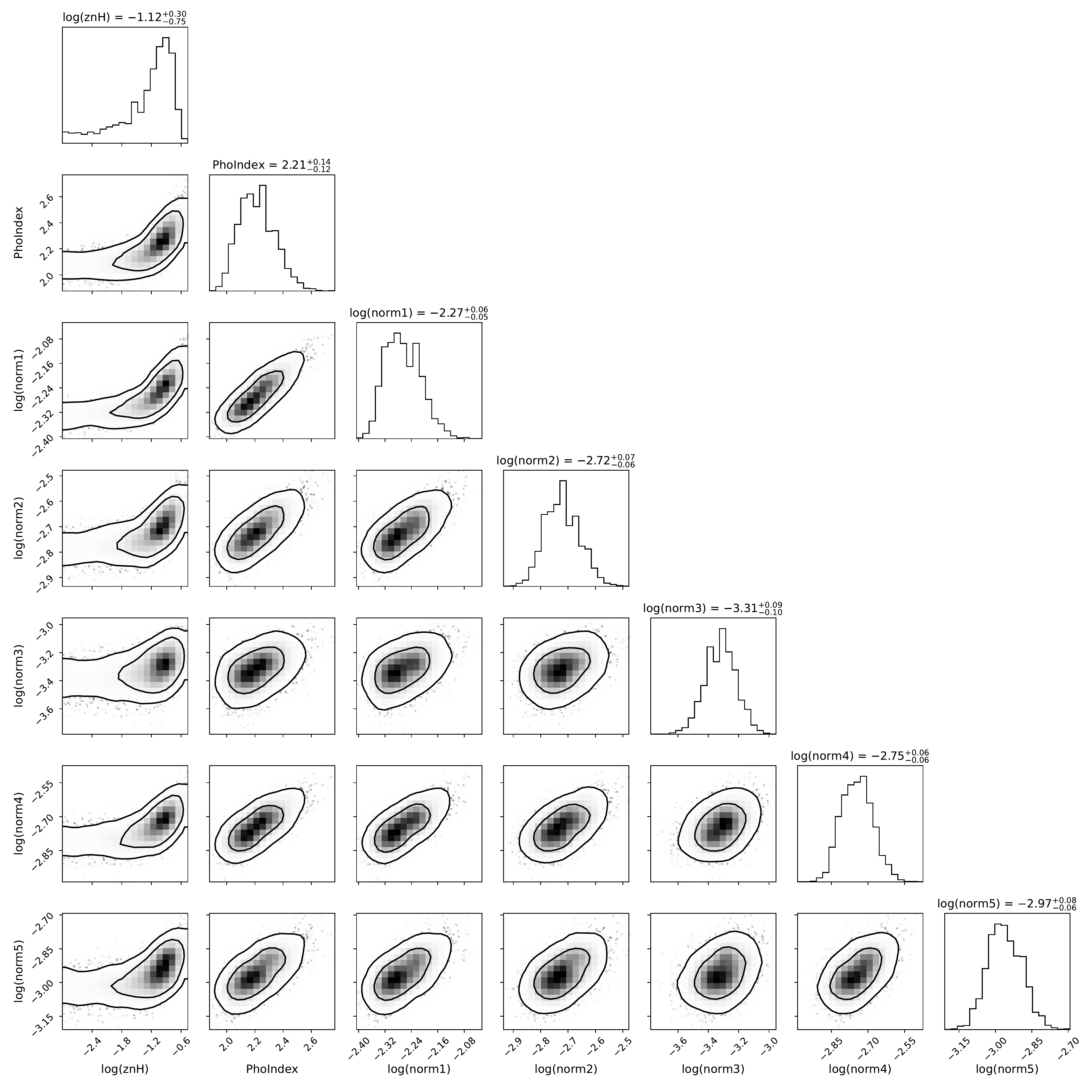}
\caption{Corner plot showing the various parameters from fitting the eROSITA spectra.}
\label{fig:cornererosita}
\end{figure*}

\begin{figure*}
\centering
\includegraphics[width=0.95\linewidth]{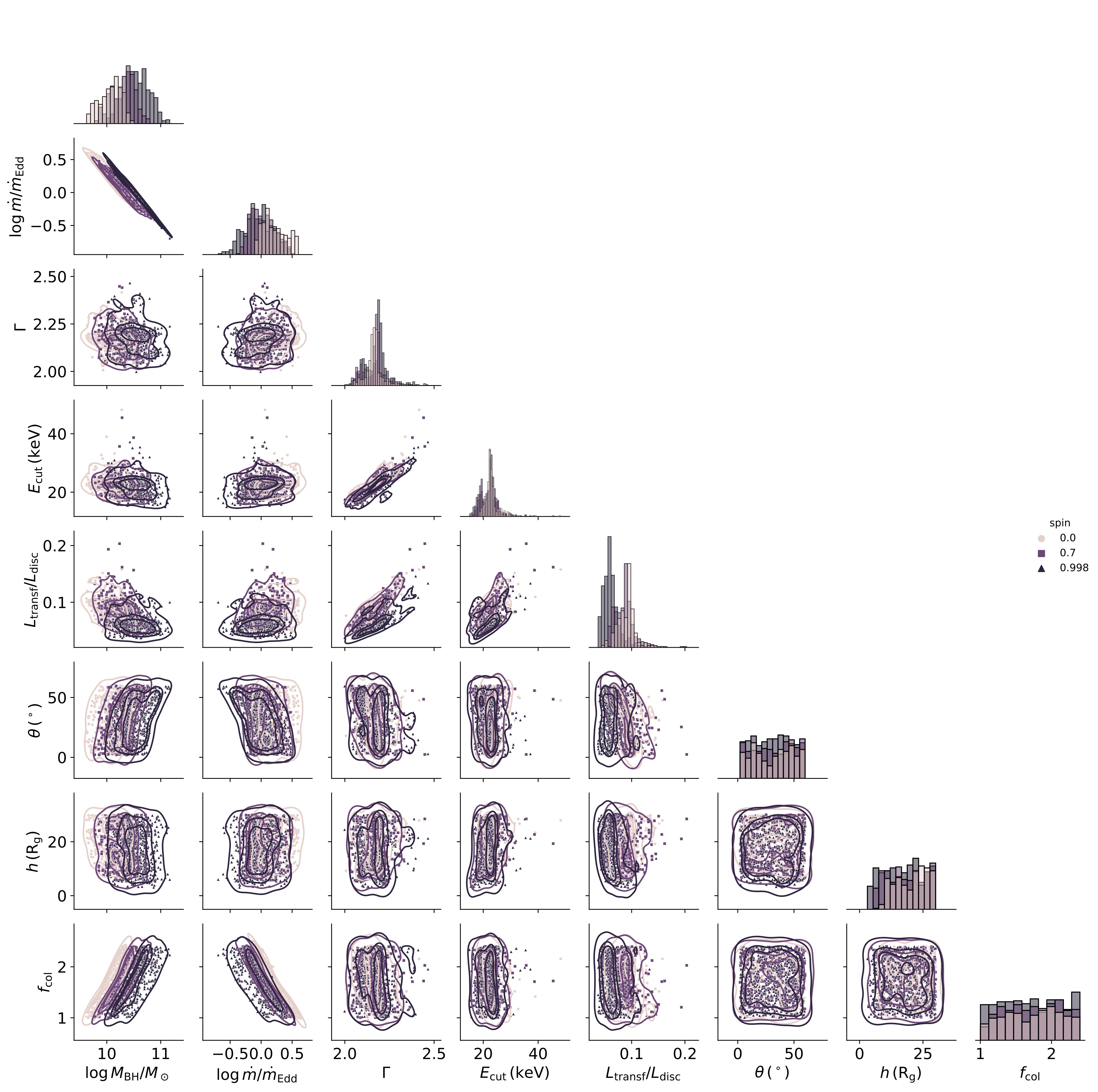}
\caption{Corner plot showing the various parameters for modelling the SED using {\tt KYNSED}. Circles, squares, and triangles correspond to spin values of 0, 0.7, and 0.998, respectively.}
\label{fig:cornerSED}
\end{figure*}

\begin{figure*}
\centering
\includegraphics[width=0.95\linewidth]{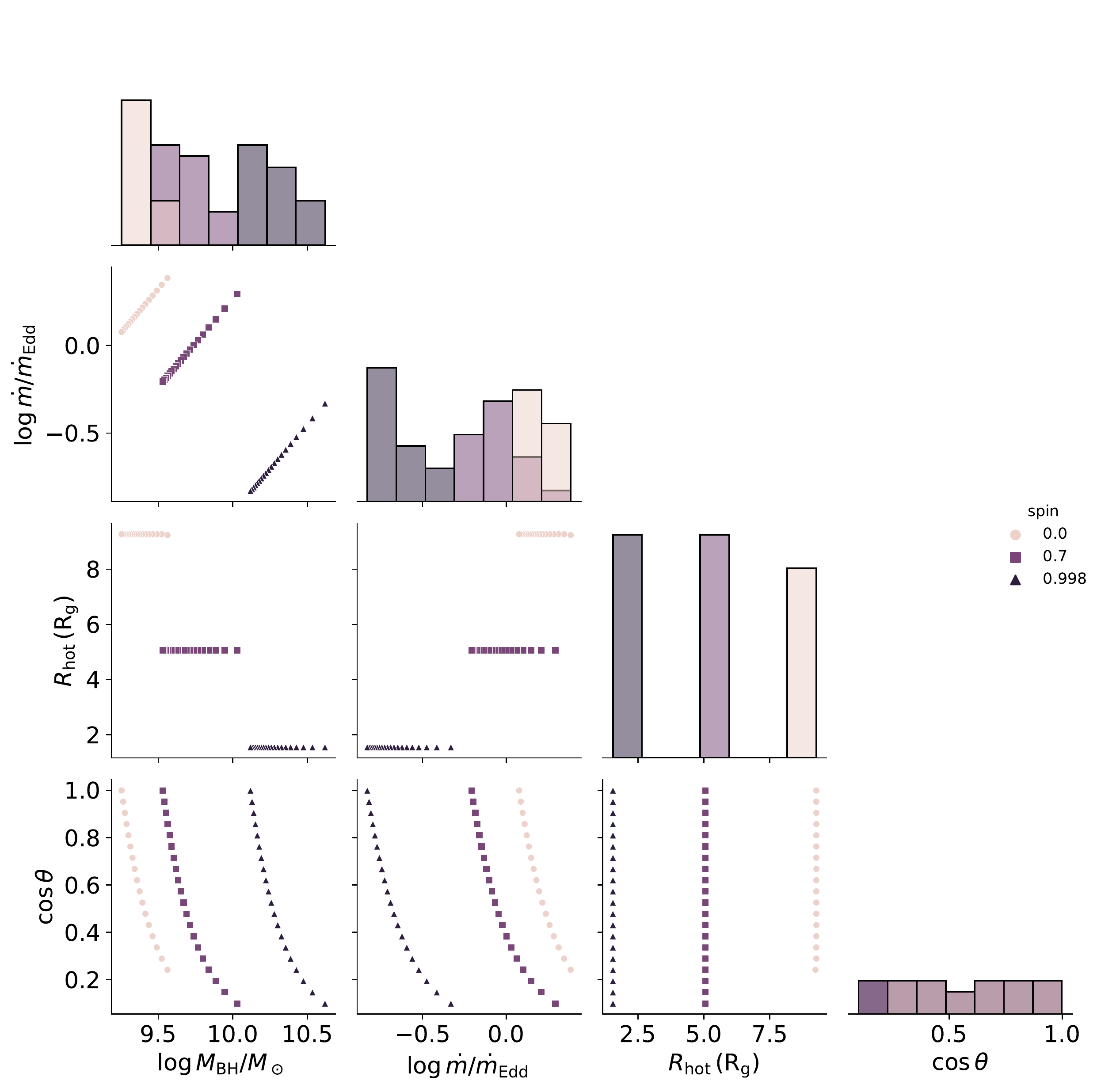}
\caption{Corner plot showing the various parameters for modelling the SED using {\tt AGNSLIM}. Circles, squares, and triangles correspond to spin values of 0, 0.7, and 0.998, respectively.}
\label{fig:cornerSED_agnslim}
\end{figure*}

Figure \ref{fig:cornererosita} shows the confidence contours obtained from fitting the eRASS spectrum as described in \S \ref{sec:erosita}. Figure \ref{fig:cornerSED} and \ref{fig:cornerSED_agnslim} show the confidence contours from modelling the SED of J1144 with {\tt KYNSED} and {\tt AGNSLIM}, respectively.

\section{Comparing KYNSED and AGNSLIM}
\label{sec:comparison}
In this section, we investigate in further detais the differences between {\tt KYNSED} and {\tt AGNSLIM} in the sub-Eddington regime. In fact, as \cite{Kubota19} mention, {\tt AGNSLIM} should be consistent with a standard accretion rate a $\mdot \lesssim 2.39$. It is only above this limit that the local flux at the emissivity peak goes above the Eddington limit (see their Fig.\,1). We consider two values of the BH spin $a^\ast=0$ and $a^\ast=0.998$. We also considered $\mdot = 0.1$ and 1. For both models, we consider a mass of $10^{10}\,\rm M_\odot$, an inclination of $0^\circ$, and $\Gamma = 2$. For {\tt KYNSED} we assume a coronal height of $10\,\rg$, $E_{\rm cut} = 50$\,keV, and $\ltransf = 0.1$. As for {\tt AGNSLIM}, we assume $kT_{\rm h} = 25\, \rm keV$, and an $R_{\rm h}$ equivalent to the radius that contains 10\%\ of the disc power \citep[as shown in Fig.\,1 of][]{Dovciak22}. This radius corresponds to $\sim 18\,\rg$ and $1.8\, \rg$, for $a^\ast=0$ and 0.998, respectively. In order to fairly compare the two models we assumed a colour correction $f_{\rm col} = 1$ for {\tt KYNSED}.


\begin{figure*}
\centering
\includegraphics[width=0.95\linewidth]{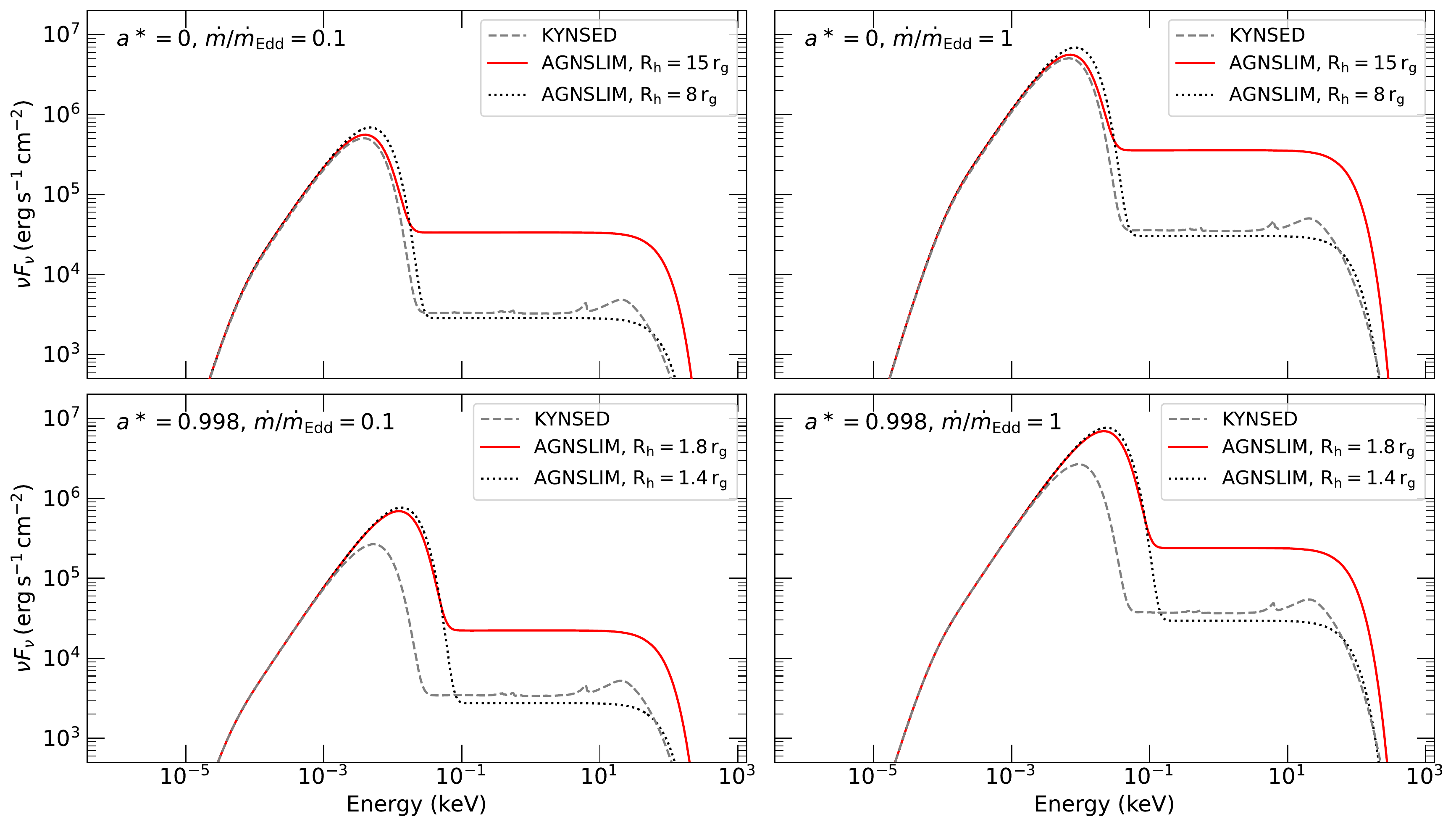}
\caption{Comparison between {\tt KYNSED} and {\tt AGNSLIM}. The top (bottom) panels show the spectra for $a^\ast = 0$ (0.998). The right and left panels show the spectra for $\mdot = 0.1$ and 1, respectively. The grey dashed spectra correspond to {\tt KYNSED}. The solid red spectra correspond to {\tt AGNSLIM} assuming a hot X-ray corona size of $R_{\rm h} = 18\,\rg$ and $1.8\,\rg$ for $a^\ast = 0$ and 0.998, respectively. The black dotted lines show the same but assuming $R_{\rm h} = 8\,\rg$ and $1.4\,\rg$ for $a^\ast = 0$ and 0.998, respectively (See Sect.\,\ref{sec:comparison} for more details). }
\label{fig:agnslim_vs_kynsed}
\end{figure*}

The resulting spectra are shown in Fig.\,\ref{fig:agnslim_vs_kynsed}. At low spin, the disc emission is slightly larger for {\tt AGNSLIM} as compared to {\tt KYNSED}. This difference increases as the spin increases. This discrepancy is most likely due to the fact that {\tt AGNSLIM} does not take general relativity (GR) effects into account. Due to GR effects, a large amount of flux from the inner disc will end up in the BH and the disc, hence the difference in the two spectra. We note that \cite{Dovciak22} found the same difference between {\tt AGNSED} \citep{Kubota18} and {\tt KYNSED} at high spin values. However, they found a different behaviour for a non-spinning BH. This is mainly due to the fact that contrary to {\tt AGNSED}, {\tt AGNSLIM} assumes that for low accretion rates, the accretion disc extends down to the ISCO, giving a better agreement with {\tt KYNSED}.

Similar to what \cite{Dovciak22} found, {\tt AGNSLIM} overestimates the X-ray emission. In order to compensate for this, we decreased the size of the X-ray region in {\tt AGNSLIM} to $8\, \rg$ and $1.4\,\rg$ for $a^\ast=0$ and 0.998, respectively. This brings the X-rays in the two models to the same level, but slightly increases the difference in the UV/optical. As discussed in Section\,3.3 in \cite{Dovciak22}, this difference arises from two factors: a) in {\tt KYNSED}, the disc flux is emitted as the cosine of the inclination while the X-rays are isotropic, and b) despite the fact that a fraction \ltransf\ is assumed in {\tt KYNSED} to be transferred to the corona, a smaller fraction in fact reaches the observer at infinity. This translates into a smaller X-ray source size in {\tt AGNSLIM}.

\bsp	
\label{lastpage}
\end{document}